\documentclass[preprint,12pt]{elsarticle}

\usepackage{epsfig}
\usepackage{graphicx}
\usepackage{xcolor}
\usepackage{amsmath}
 \usepackage{booktabs} 
\DeclareMathAlphabet      {\mathsf}{OT1}{cmss}{m}{n}
\SetMathAlphabet\mathsf{normal}{OT1}{cmss}{m}{n}
\DeclareFontFamily{U}{musix}{}%
\DeclareFontShape{U}{musix}{m}{n}{%
  <-12>   musix11
  <12-15> musix13
  <15-18> musix16
  <18-23> musix20
  <23->   musix29
}{}%
\newcommand*\musix{\usefont{U}{musix}{m}{n}\selectfont}
\DeclareTextFontCommand{\textmusix}{\musix}

\usepackage{amssymb}
\usepackage{amsthm}
 \usepackage{subfigure}
\usepackage{lscape}
\usepackage{caption}
\journal{Chaos, Solitons \& Fractals}

\begin{document}

\begin{frontmatter}



\title{Higher-Order Network Representation of J. S. Bach’s Solo Violin Sonatas and Partitas: Topological and Geometrical Explorations} %% Article title


\author[inst1,inst2]{Dima Mrad}
\author[inst1,inst2]{Sara Najem}
\affiliation[inst1]{Department of Physics, American University of Beirut, Beirut, Lebanon}
\affiliation[inst2]{Center for Advanced Mathematical Sciences, American University of Beirut, Beirut, Lebanon}

%% Abstract
\begin{abstract}
Music is inherently complex, with structures and interactions that unfold across multiple layers. Complex networks have emerged as powerful structures for the quantitative analysis of Western classical music, revealing significant features of its harmonic and structural organization. Although notable works have used these approaches to study music, dyadic representations of interactions fall short in conveying the underlying complexity and depth. In recent years, the limitations of traditional graph representations have been questioned and challenged in the context of interactions that could be higher-dimensional. Effective musical analysis requires models that capture higher-order interactions and a framework that simultaneously captures transitions between them. Subsequently, in this paper, we present a topological framework for analyzing J. S. Bach's Solo Violin Sonatas and Partitas that uses higher-order networks where single notes are vertices, two-note chords are edges, three-notes are triangles, etc. We subsequently account for the flow of music, by modeling transitions between successive notes. We identify genre-specific patterns in the works' geometric and topological properties. In particular, we find signatures in the trends of the evolution of the Euler characteristic and curvature, as well as examining adherence to the Gauss-Bonnet theorem across different movement types. The distinctions are revealed between slow movements, Fugues, and Baroque dance movements through their simplicial complex representation.
\end{abstract}

\begin{graphicalabstract}
\begin{figure}[h]
    \centering
    \includegraphics[width=0.7\linewidth]{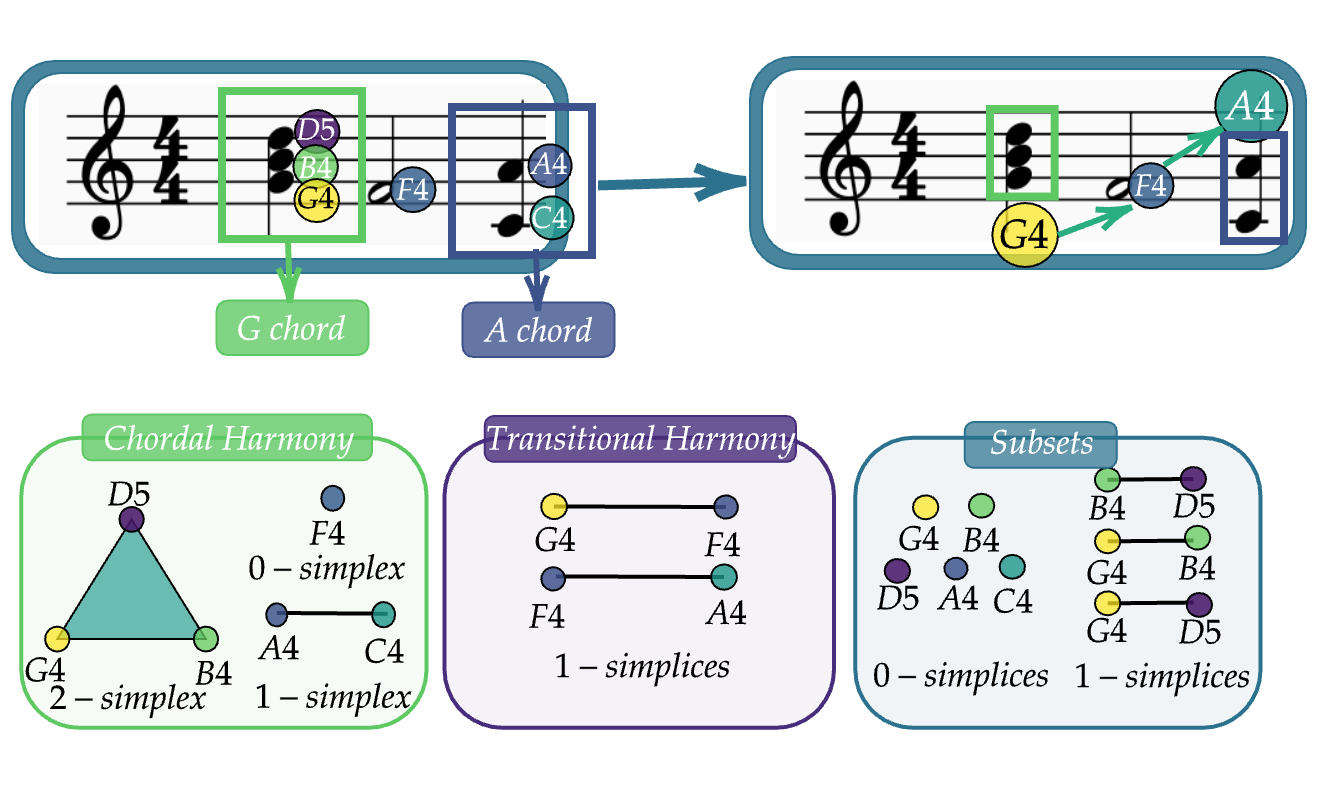}
\end{figure}
\end{graphicalabstract}

\begin{highlights}
\item We provide a framework for representing musical pieces using higher-order networks.
\item In this framework, simplicial complexes model harmonic-melodic interactions.
% \item Simplicial complexes allow the study of interactions between harmony and melody. 
\item Their topological measures, such as the Betti numbers, are used to classify pieces.
\item Additionally, geometrical notions like curvature are computed for classification. 
\item Gauss-Bonnet-like relation between their geometry and topology is presented. 
\end{highlights}

%% Keywords

\end{frontmatter}



\section{Introduction}
\label{sec:introduction} 
In recent years, mathematical approaches to musical analysis have increasingly revealed their proficiency in capturing structural and organizational features of music that traditional qualitative methods fail to depict \cite{liu2013statistical,berezovsky2019structure,Meredith}. 
The framework of network science opened new frontiers in music research, offering robust tools to model complex musical relationships. These network-theoretic approaches have become prevalent in analyzing interactions between the core structural elements of music, encompassing pitch, rhythm, and harmony \cite{xin2016complex,ferretti2018complex,gomez2014complex,kulkarni2024information,gomez2025network,di2025decoding,chi2008analyzing,tsai2024depth}.
Johann Sebastian Bach composed his Six Sonatas and Partitas for Solo Violin BWV $1001-1006$ in the $17^{\text{th}}$ century, between $1703$ and $1720$ during his time in Cöthen where he was serving as Kapellmeister to Prince Leopold. In the 17th century, the violin and other da braccio (by arm) instruments were still developing their independent repertoire. As a result, compositions for solo violin (without accompaniment) were not common, and most violin compositions depended on figured bass. Solo violin music was often improvised, and shared through oral tradition. Following the mid-17th century, few works have been preserved, including Bach’s unaccompanied violin compositions. These Sonatas and Partitas stand out as one of the most remarkable and inventive contributions to the violin repertoire, giving a profound insight into the instrument’s technical skill and expressive power. The work is made up of six compositions in total, three sonatas and three partitas. Johann Sebastian Bach’s Solo Violin Sonatas and Partitas, comprise six violin pieces known for their rich structures and dynamics with combined melody, harmony, and polyphony \cite{moosbauer2017johann,lester1999bach}.

\noindent Although notable existing works have yielded valuable insights in the context of using network models to study Bach's compositions \cite{kulkarni2024information,gomez2014complex}, their reliance on pairwise interactions presents a limitation in the study of musical relationships. These approaches fail to record essential features of polyphonic music encoded in higher-order interactions, like simultaneously played notes and chords.

\noindent To account for such interactions, researchers have sought ways to include higher-order structures in analytical frameworks and some measures have been introduced, for instance, the local O-information approach \cite{scagliarini2022quantifying} which analyzes the interactions of chords in Bach's chorals and captures different levels of redundancy and statistical synergy in chords across the musical piece.

\noindent Subsequently, the challenge lies in developing a representation that goes beyond traditional models and can encode these higher-order complexities. In this paper, we construct musical pieces as simplicial complexes, which are structures made up of $k-$simplices which are sets of  $k+1$ nodes. 
In this representation, a 0-simplex is a node, which can be interpreted as an individual note in a musical context, a 1-simplex corresponds to an edge or 2-note chord, a 2-simplex is a set of 3 nodes constituting a filled triangle or a 3-note chord, and higher-dimensional simplices correspond to chords with more notes played simultaneously. The chords in this context are defined to be a set of musical notes played simultaneously, whether by a solo instrument or a group of instruments. This provides a representation for vertical elements of the chordal structure. Then, we account for the horizontal correlations between the notes by considering only the nearest neighbors in the musical line, forming transition edges between them. Next, we proceed by computing the boundary operators, which generalize the incidence matrices used in graph theory. %The boundary operator \( \partial_n \) can be represented by matrices \( B_{[k]} \), where \( B_{[k]}(i, j) \) indicates if an \( (k-1) \)-simplex is part of the boundary of an \( k \)-simplex. The boundary matrix \( B_{[k]} \) is of size \( N_{[k-1]} \times N_{[k]} \), with \( N_{[k]} \) representing the number of \( k\)-simplices in the complex. 
This incidence matrix encodes the relationships between the simplices. The higher-order Laplacian of a simplicial complex is then recovered from the incidence matrix. In regular graphs, the Laplacian is a fundamental operator that describes diffusion from a node to another node through links, the higher-order Hodge Laplacian is a generalization that describes diffusion from an $k$-simplex to another $k$-simplex. An important property that arises from the Hodge decomposition of these higher-order Laplacians is that the degeneracy of the zeroth eigenvalues of the $k$-Laplacian equals the Betti number $\beta_k$. These characterize the number of connected components, like in traditional networks, as well as their higher-order cavities. The Euler characteristic is then calculated using the obtained Betti numbers, it offers a single numerical descriptor of the topology, and represents a global topological invariant that captures the overall shape of a simplicial complex. Following this, we transition from topology to geometry and define the discrete curvature measure on simplicial complexes: the Forman-Ricci curvature, and evaluate it at different orders of the simplicial complex. In principle, we are interested in analyzing the musical data as it evolves temporally, for this propose we propose the implementation of two main approaches: The first, which we call \textit{the cumulative approach}, reconstructs the piece as a progressive temporal evolution, starting with the first measure and sequentially adding others until the full composition takes shape. This approach reflects the composer’s perspective, revealing the structure as it develops and grows over time. The second approach, which is a sliding window approach, was employed in selected scenarios, and it comprises systematically shifting the window across the piece providing a dynamic, time-sensitive representation of musical features. This approach captures the piece instantaneously as a listener would perceive it in real time; it hence represents the \textit{listener's perspective}. In the implementation of these two approaches, we succeeded in establishing a correspondence between topological metrics and musical form, which refers to the overall structure and organization of a musical composition, describing how sections of a piece are arranged and related to one another. 
For the analysis, we consider the Sonatas and Partitas for violin solo by J. S. Bach, composed in the late 1710s. There are three four-movement sonatas in G
minor, A minor and C major, and three partitas in B minor, D minor and E major. The works have been of particular interest to music analysts, as they comprise a stylistically coherent set
of works of the highest musical quality. For comparative purposes, we also consider a set of works by Max Reger, Ysaye, Vaclav Pichl, etc.

\section{Structure of Higher-Order Networks}
\label{subsec:higherordernets}
\subsubsection{Higher-order networks}
\label{subsubsec:higherordernets}
\begin{figure}[!htp]
    \centering
    \includegraphics[width=0.7\linewidth]{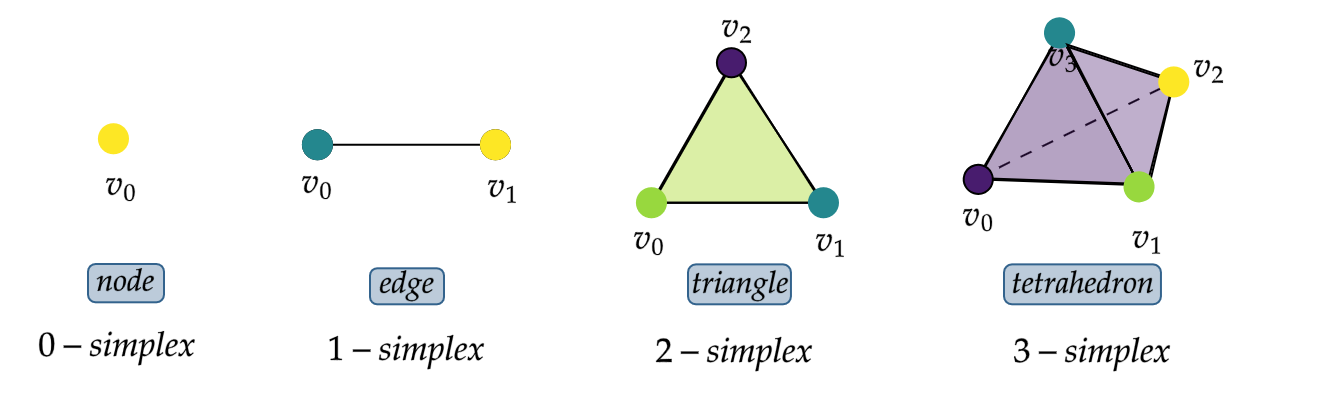}
    \caption{Illustration showing simplices of different orders: a 0-simplex (node), 1-simplex (edge), 2-simplex (triangle), and 3-simplex (tetrahedron).}
    \label{simplices}
\end{figure}

A higher order network $\mathcal{N} = (V, H) $ is a set of vertices $V$ and hyperedges $H$ where any element of $H$ serves as a single hyperedge. Simplicial complexes are one example of higher-order networks, where higher-order interactions are represented as simplices. A simplex of dimension $k$ is a set of $k+1$ nodes,$\alpha = \left[v_0, v_1, v_2, ..., v_k\right]$ which capture the interactions between the nodes (Figure \ref{simplices}), these higher-order interactions are distinguished from those in regular graphs by considering simultaneous interactions between the nodes. For instance, a triangle formed of nodes $v_0, ~v_1, ~v_2$ represents simultaneous interactions between the nodes as opposed to pairwise interactions of an empty triangle in a regular graph. This generalizes to higher-order simplices.  Simplicial complexes are built out of sets of interacting nodes of any dimension, and are closed under the inclusion of subsets by definition. \cite[Chapter~2]{bianconi2021higher}. 

\subsection{Oriented simplices}
\begin{figure}[h]
    \centering
    \includegraphics[width=0.8\linewidth]{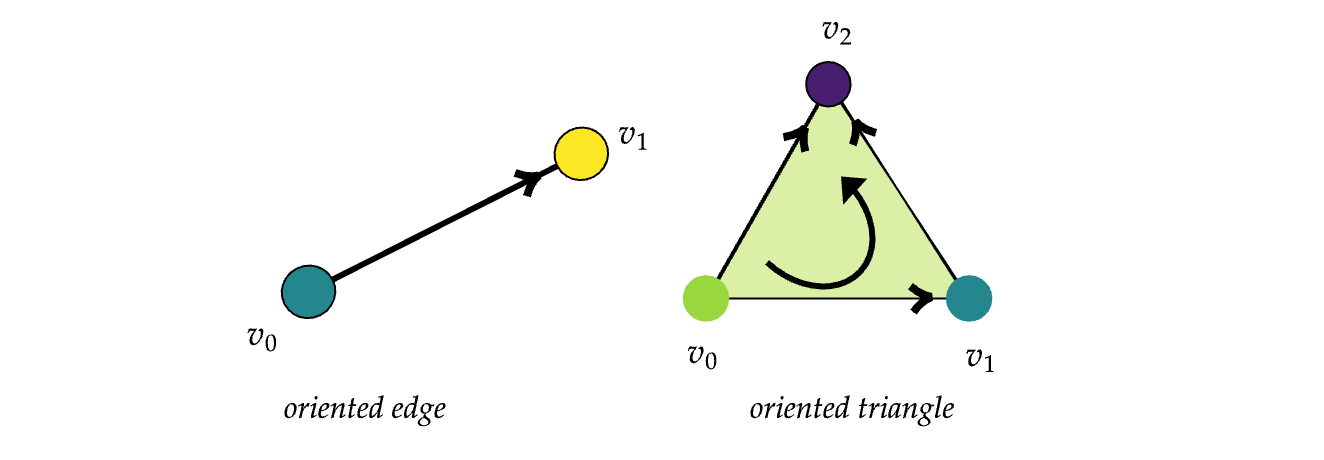}
    \caption{Illustration of oriented simplices: an oriented $1-$simplex (edge) and an oriented $2-$simplex (triangle).}
    \label{fig:orientedSimplices}
\end{figure}

An $n-$dimensional oriented simplex $s = [i_0, i_1, i_2, \cdots i_n]$ is a set of $(n+1)$ nodes associated with an orientation defined to be $[i_0, i_1, i_2, \cdots i_n] = (-1)^{\sigma(\pi)} [i_{\pi(0)}, i_{\pi(1)}, \cdots, i_{\pi(n)}]$ with $\sigma(\pi)$ indicator of the parity of the orientation, this is illustrated in Fig. \ref{fig:orientedSimplices} \cite{bianconi2021higher,millan2022geometry,torres2020simplicial}.

\subsection{Boundary operator and Incidence matrices}
\begin{figure}[!t]
    \centering
    \includegraphics[width=0.8\linewidth]{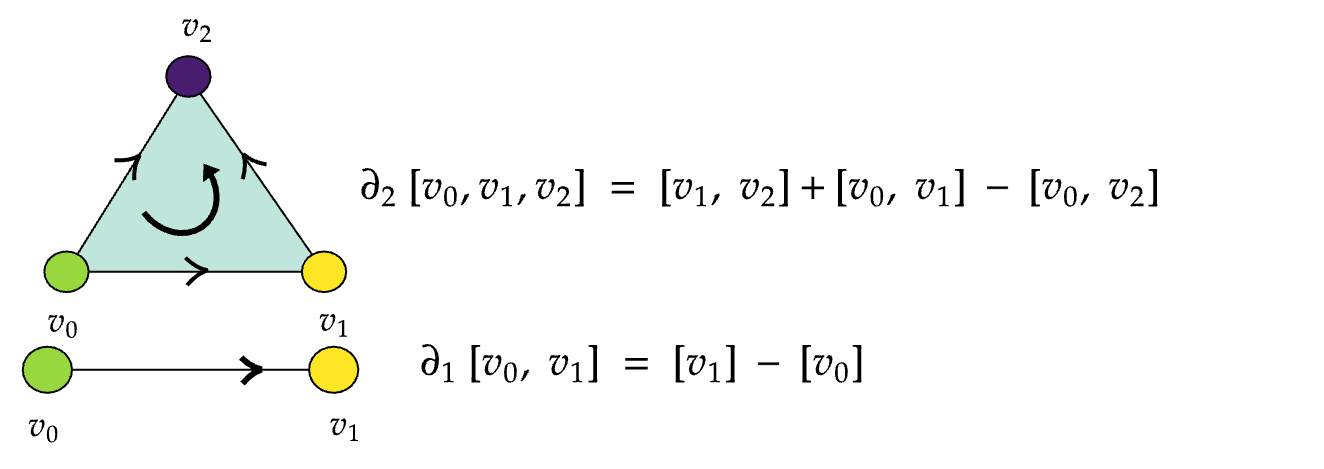}
    \caption{Illustration of the boundary operator \( \partial \)'s action on a simplex, demonstrating how it maps an \( n \)-simplex to an \( (n-1) \)-chain of its boundary simplices.}
    \label{BoundaryMap}
\end{figure}

The boundary operator denoted as \( \partial_n \), operates on chains within the free abelian group \( C_n \). In this case, an \( n \)-chain \( \sigma \) is a linear combination of \( n \)-simplices \( \alpha \), where each simplex \( \alpha \) is represented with integer coefficients \( z_{\alpha} \): $\sigma = \sum_{\alpha} z_{\alpha} \alpha$. The boundary operator \( \partial_n \) maps an \( n \)-simplex to an \( (n-1) \)-chain as follows $\partial_n[v_0, v_1, \ldots, v_n] = \sum_{p=0}^{n} (-1)^p [v_0, \ldots, \widehat{v_p}, \ldots, v_n]$, where \( [v_0, \ldots, \widehat{v_p}, \ldots, v_n] \) denotes the simplex with node \( v_p \) omitted. The action of the boundary operator on a simplex is illustrated in Figure \ref{BoundaryMap} \cite{bianconi2021higher,millan2022geometry,torres2020simplicial}.

\noindent In the context of higher-order networks, this operator generalizes incidence matrices used in graph theory. It can be represented by matrices \( B_{[n]} \), where \( B_{[n]}(i, j) \) indicates if an \( (n-1) \)-simplex is part of the boundary of an \( n \)-simplex. Specifically, \( B_{[n]} \) is of size \( N_{[n-1]} \times N_{[n]} \), with \( N_{[n]} \) representing the number of \( n \)-simplices in the complex \( \mathcal{K} \), and it satisfies the property \( B_{[n]} B_{[n+1]} = 0 \). The entries of the incidence matrices are given by binary values indicating the inclusion of lower-dimensional simplices in higher-dimensional ones and are computed using $\left[B_{[m]}\right]_{\alpha ', \alpha} =  (-1)^p$, for every pair of $m$- dimensional simplex $\alpha$ and $(m-1)$- dimensional simplex $\alpha '$ such that: $\alpha = [v_0, v_1, ..., v_m];\alpha' = [v_0, v_1, ..., v_{p-1}, v_{p+1}, ..., v_m]$ \cite[Chapter~3]{bianconi2021higher}. Alternatively, the components can be expressed more formally as given in Equation \ref{EqIncidence} \cite{bianconi2021higher,millan2022geometry,torres2020simplicial,millan2025topology}.

\begin{equation}\label{EqIncidence}
B_{k_{\alpha',\alpha}} =
\begin{cases}
+1 & \text{if } \alpha' \subset \alpha \text{ and the orientations align}, \\
-1 & \text{if } \alpha' \subset \alpha \text{ and the orientations do not align}, \\
0 & \text{if } \alpha' \nsubseteq \alpha.
\end{cases}
\end{equation}

\begin{figure}[!t]
    \centering
    \includegraphics[width=0.8\linewidth]{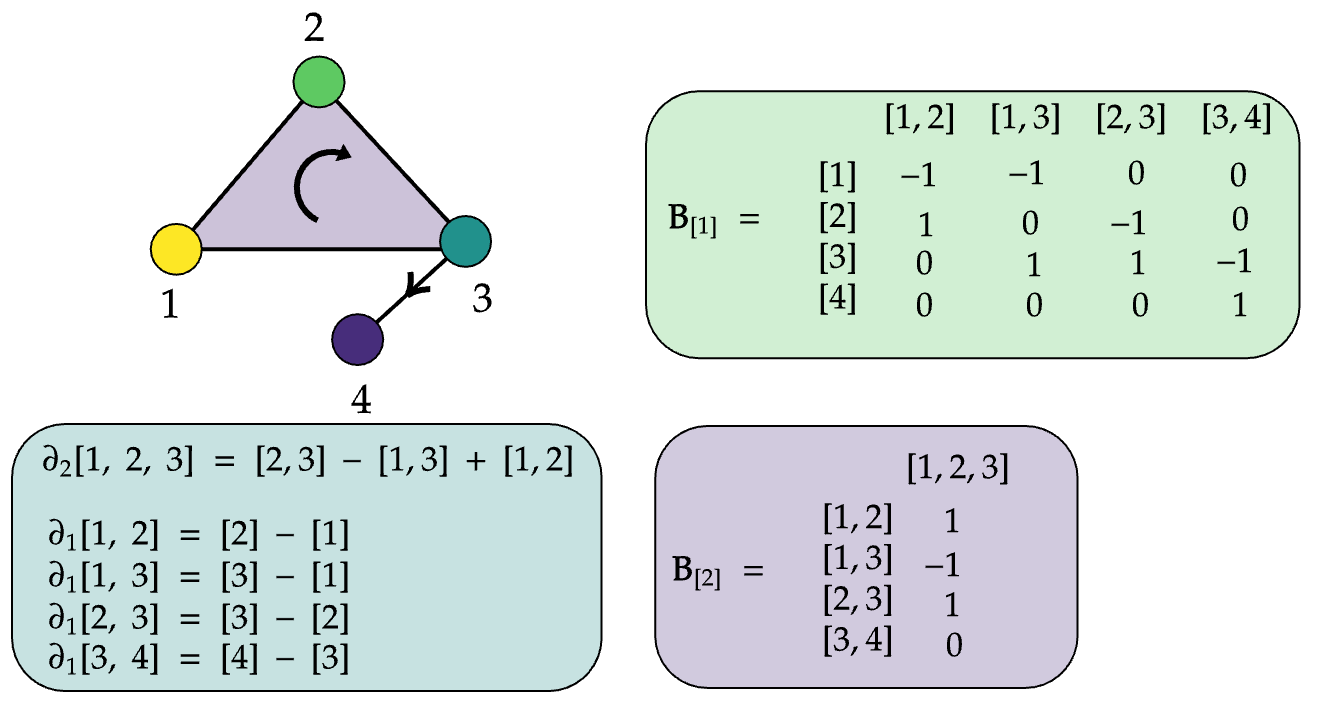}
    \caption{Example demonstrating the computation of incidence matrices $B_1$ and $B_2$ for a simplicial complex of dimension $d = 2$}
    \label{IncidenceMat}
\end{figure}

An example of computing incidence matrices for a simplicial complex is shown in Figure \ref{IncidenceMat} . Boundary matrices possess the property that the boundary of the boundary is always null, which can be expressed mathematically as \
\begin{equation}\label{eq:boundarynilpotency}
B_{k} B_{k+1} = 0 
\end{equation} 

\subsection{Higher-order Hodge Laplacian}
The traditional graph Laplacian defined as $L = D - A$ is a fundamental operator employed in the study of dynamics on a network, it can be alternatively written as
$L = BB^T$, where $B$ is the Boundary (Incidence) matrix. For a $k$-simplicial complex, the $n$-order Hodge Laplacian is represented by the matrix:
\begin{equation}\label{eq:Laplacian}
	L_{[n]} = B_{[n]}^T B_{[n]} + B_{[n+1]} B_{[n+1]}^T
\end{equation}
 This operator describes diffusion between $k$-simplices: The second term describes diffusion between $k$-simplices through shared $(k+1)$-simplices, is denoted $L_{[k]}^{up}$, and represents a higher-order analogue of diffusion on a network through shared edges. The first term, denoted $L_{[k]}^{down}$ does not exist on networks, and describes diffusion through shared $(k-1)$-simplices. \cite[Chapter~3]{bianconi2021higher}.

\section{Topology on Higher-order networks}
\subsection{Homology groups and Betti numbers}\label{subsec:homologyGrp}

\begin{figure}
    \centering
    \includegraphics[width=0.8\linewidth]{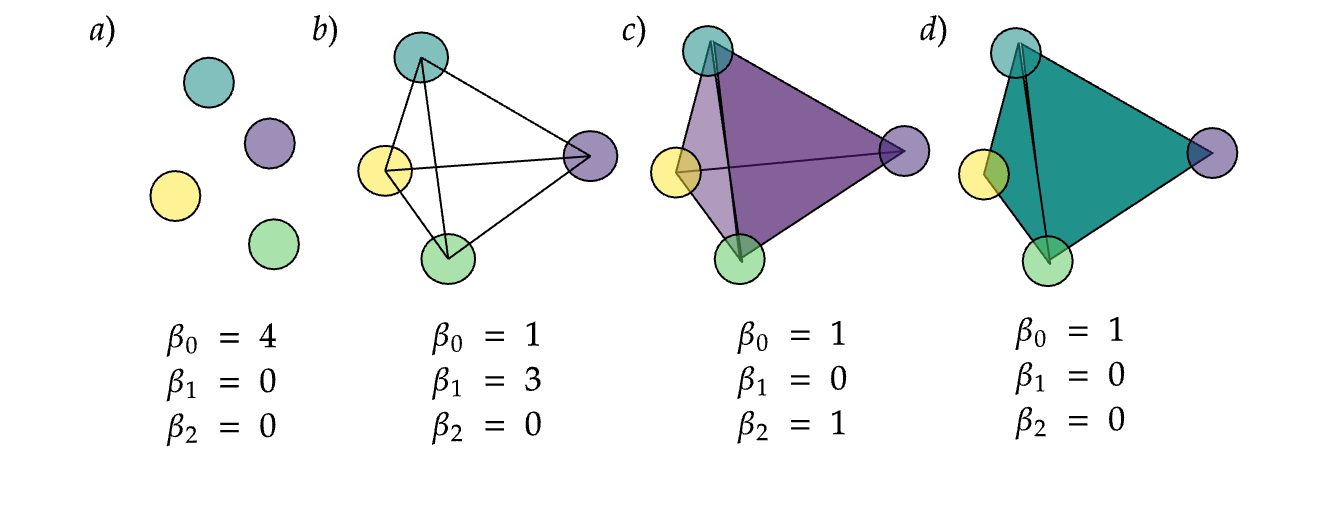}
    \caption{Examples for Betti numbers on simplicial complexes. In panel $a)$, the simplicial complex is composed of 4 vertices, the corresponding non-zero betti number is $\beta_0 = 4$ which represents the number of independent connected components (in this case, each of the vertices is one connected component). In panel $b)$, the simplicial complex is composed of 4 vertices connected by 6 edges forming 3 2D triangles corresponding to 3 $1-$dimensional holes, and 1 connected component. The simplicial complex of Panel $c)$ is formed of 4 filled triangles, creating a  a \textit{hollow} tetrahedron with a 2D void inside. Finally, panel $d)$ is a filled tetrahedron, it consists of an \textit{interaction} of dimension 4, forming one solid tetrahedron as one connected component.}
    \label{fig:betti}
\end{figure}

The $k^{th}$ homology group $\mathcal{H}_k$ is defined to be the quotient space:
$$\mathcal{H}_k = \frac{\ker(\partial_k)}{\text{im}(\partial_{k+1})}$$
The $k^{th}$ homology group represents $k-$cycles that do not form the boundary of any $(k+1)-$chain. 
In other words, $\ker(\partial_k)$ consists of $k-$cycles which are chains with zero boundary (they form closed loops on the level of dimension $k$). On the other hand, $\text{im}(\partial_{k+1})$ consists of $(k+1)-$boundaries which are chains that can be written as the boundary of some $(k+1)-$dimensional chain \cite{munkres2018elements,Hatcher:478079,bianconi2021higher}. 
The quotient space gives the difference between cycles that are in the kernel of $\partial_m$ and those that correspond to the boundaries of higher-dimensional somplices yielding the true $k-$dimensional independent holes in the simplicial complex. These independent holes of dimension $k $ are topological invariants and are called the \textbf{\textit{Betti numbers}} of order $k$, an illustration is provided in Figure \ref{fig:betti}.

\subsection{Kernel of the higher-order Hodge Laplacian}
The degeneracy of the zero eigenvalue of the Hodge-Laplacians corresponds to the number of connected components for the $L_{[0]}$ network Laplacian in classical spectral graph theory, and generalizes to the number of $k+1$ dimensional voids or $k$-cavities for each $L_{[k]}$ \cite{bianconi2021higher}. For instance, for the $L_{[1]}$, the betti number captures the number of unfilled loops or $1$-dimensional voids in the simplicial complex, for $L_{[2]}$, we obtain the number of $2$-dimensional voids and generalizing to $L_{[m]}$, the degeneracy of the zero eigenvalue indicates the presence of higher-dimensional holes that are not filled by higher-dimensional simplices. We provide a proof for the above statement in what follows:
the $\mathbf{L_k^{\text{up}}} $ term of the Hodge Laplacian describes diffusion between simplices of order $k$ and shared simplices of order $(k+1)$, while the $\mathbf{L_k^{\text{down}}}$ term describes diffusion between simplices of order $k$ and shared simplices of order $(k-1)$ \cite{bianconi2021higher}.\\
Recalling the property of the boundary matrices given in Equation \ref{eq:boundarynilpotency}, we can write:
\begin{equation} \label{kerLup}
\begin{aligned}
\mathbf{L}_k^{\text{up}} \mathbf{L}_k^{\text{down}} &= \mathbf{B}_{k+1} \mathbf{B}_{k+1}^{\text{T}} \mathbf{B}_k^{\text{T}} \mathbf{B}_k \\
&= \mathbf{B}_{k+1} [\mathbf{B}_k \mathbf{B}_{k+1}]^{\text{T}} \mathbf{B}_k \\
&= 0
\end{aligned}
\end{equation}

\begin{equation}\label{kerLdown}
\mathbf{L}_k^{\text{down}} \mathbf{L}_k^{\text{up}} = \mathbf{B}_k^{\text{T}} \mathbf{B}_k \mathbf{B}_{k+1} \mathbf{B}_{k+1}^{\text{T}}
\end{equation}

Equations \ref{kerLup} and \ref{kerLdown} imply that the commutator relations  are as given by: 

\begin{equation}\label{Lcommutation}
\begin{aligned}
\left[ \mathbf{L_k^{\text{up}}}, \mathbf{L_k^{\text{down}}} \right] &= \mathbf{L_k^{\text{up}} L_k^{\text{down}}} - \mathbf{L_k^{\text{down}} L_k^{\text{up}}} = 0 \\
\left[ \mathbf{L_k^{\text{down}}}, \mathbf{L_k^{\text{up}}} \right] &= \mathbf{L_k^{\text{down}} L_k^{\text{up}}} - \mathbf{L_k^{\text{up}} L_k^{\text{down}}} = 0 \\
\left[ \mathbf{L_k^{\text{down}}}, \mathbf{L_k} \right] &= 0 \\
\left[ \mathbf{L_k^{\text{up}}}, \mathbf{L_k} \right] &= 0
\end{aligned}
\end{equation}

As a result of Equation \ref{Lcommutation}, the three operators $\mathbf{L_k^{\text{down}}}$, $\mathbf{L_k^{\text{up}}}$ and $\mathbf{L_k}$ can be simultaneously diagonalized, their corresponding eigenvectors obey the following relations:

\[ \text{im}(\mathbf{L^{{\text{down}}}_{k}}) \subseteq \ker(\mathbf{L^{\text{{up}}}_{k}}), \]
\[ \text{im}(\mathbf{L^{{\text{up}}}_{k}}) \subseteq \ker(\mathbf{L^{{\text{down}}}_{k}}). \]
$$\ker(\mathbf{L_k}) = \ker(\mathbf{L^{\text{up}}_k})
\cap \ker(\mathbf{L^{\text{down}}_k}).
$$
The eigenvectors of \( \mathbf{L}^{k} \) are therefore classified into the following categories:  
\begin{itemize}
    \item[-] An eigenvector is a simultaneous eigenvector of \( \mathbf{L}^{\text{down}}_{k} \) and $\mathbf{L}^{\text{up}}_{k}$ corresponding to a nonzero eigenvalue \( \lambda^{k} \) of $\mathbf{L}^{\text{down}}_{k}$ , and a zero eigenvalue of \( \mathbf{L}^{\text{up}}_{k} \).
    \item[-] An eigenvector is a simultaneous eigenvector of \( \mathbf{L}^{\text{down}}_{k} \) and $\mathbf{L}^{\text{up}}_{k}$ corresponding to a nonzero eigenvalue \( \lambda^{k} \) of $\mathbf{L}^{\text{up}}_{k}$ , and a zero eigenvalue of \( \mathbf{L}^{\text{down}}_{k} \).
    \item[-] An eigenvector is an eigenvector common to \( \mathbf{L}_{k} \), \( \mathbf{L}^{\text{down}}_{k} \), and \( \mathbf{L}^{\text{up}}_{k} \), corresponding to the eigenvalue \( \lambda^{k} = 0 \).
\end{itemize}

The \textbf{harmonic} space is given by: $\ker(\mathbf{L_k}) = \ker \mathbf{L}_k^{up} \cap \ker \mathbf{L}_k^{down} $. So we can write:
\begin{equation}\label{bettiproof}
\begin{aligned}
    \text{dim} \ker(\mathbf{L_k}) &= \text{dim} \ker \mathbf{L}_k^{up} \cap \mathbf{L}_k^{down}  \\
    &= \text{dim} \ker \mathbf{L}_k^{down} - \text{dim} \, \text{im} \, \mathbf{L}_k^{up}  \\
    &= \text{dim} \, \text{im} \, \mathbf{L}_k^{down} - \text{dim} \ker \, \mathbf{L}_k^{up}  \\
    &= \text{dim} \ker \mathbf{B}_k - \text{dim} \, \text{im} \, \mathbf{B}_{k+1} \\
    &= \text{rank} \ \mathcal{H}_k \\
    &= \beta_k
\end{aligned}
\end{equation}

\subsection{Topological invariants in simplicial complexes}

The Betti numbers make up an important topological invariant that describes the connectivity and higher-dimensional structure of simplicial complexes. Another topological invariant can be obtained from the Betti numbers. According to the Euler–Poincaré formula, the Euler characteristic $\chi$ of a simplicial complex can be expressed in terms of the Betti numbers as:
\begin{equation} \label{eq:EulerChar}
	 \chi = \sum_{m \geq 0} (-1)^m \beta_m
\end{equation}

\section{Geometry on simplicial complexes}
Geometric notions such as curvature help describe the local shape of networks, how tightly connected or stretched different parts are. Definitions of curvature can be either continuous Ricci scalar curvature in Riemannian geometry, or discretized versions of curvature like the Gaussian, Mean and Regge curvatures, both of which necessitate properly defined  notions of angle and length. In our context, networks (and higher-order networks) are not geometrical, objects, they are abstract topological structures with no equivalent of angle and length. Thus, defining a notion curvature over these structures requires special treatment \cite{boguna2021network,krioukov2010hyperbolic,kitsak2017latent,mulder2018network,wu2015emergent,bianconi2017emergent}. One convenient notion is Forman curvature, a discrete analogue of Ricci curvature, which can be applied  to graphs as well as higher-dimensional complexes.

\subsection{The Combinatorial Bochner-Weitzenböck identity}
In Riemannian geometry, the Bochner–Weitzenböck identity decomposes the Hodge Laplacian on 
$p-forms$ into a sum of a nonnegative second-order differential operator and a zeroth-order curvature term. In Ref. \cite{forman2003bochner}, the combinatorial analog is introduced using the Hodge Laplacian of order $k$, which is decomposed into a nonnegative matrix operator that plays the role of a discrete rough Laplacian $B$ and a diagonal matrix that can be interpreted as a combinatorial curvature term $F$.
\begin{equation}
	L_k = B_k + F_k
\end{equation} 

%\subsection{Forman curvature on unweighted graphs}
In the context of unweighted graphs, the Forman curvature is a scalar value assigned to each edge $e = (u, v)$ based on the degrees of the nodes it connects \cite{saucan2021simple,weber2017characterizing,roy2020forman,sreejith2016forman}. For an unweighted graph, it is given by:
\begin{equation} \label{unweightedSimpleForman}
F(e) = 4 - (\text{deg}(u) + \text{deg}(v))
\end{equation}
Equation \ref{unweightedSimpleForman} indicates that edges connecting high-degree nodes have negative curvature, while edges within densely connected clusters tend to have positive curvature.

\section{Methodology}
\subsection{Construction}
\begin{figure}[h]
    \centering
    \includegraphics[width=0.8\linewidth]{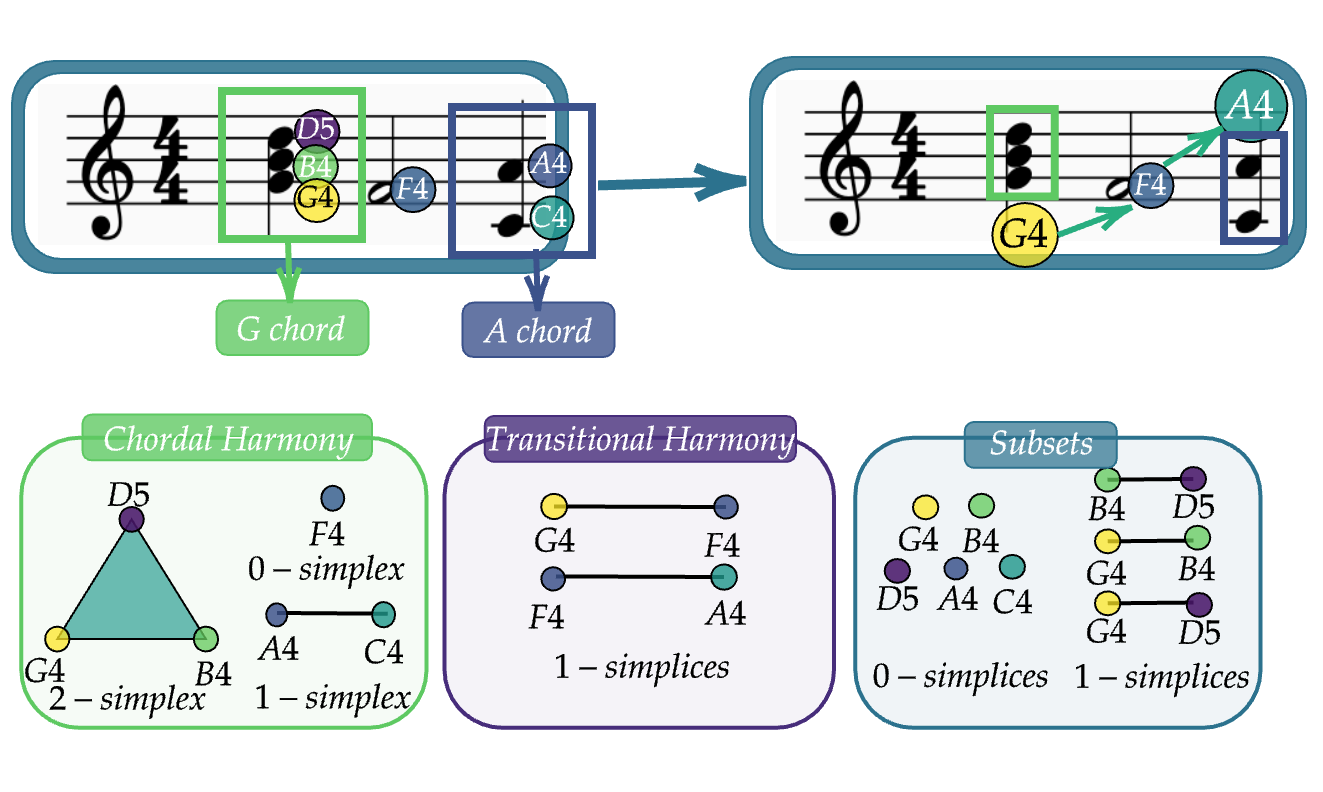}
    \caption{The simplices of different order for a given line of music}
    \label{musicSimplices}
    % \label{fig:kmeans-comparison}
\end{figure}

We begin our analysis by reading the MIDI file associated with the movement using the MIDI Toolbox \cite{eerola2004midi},
which generates a matrix, $M$. The number of rows in $M$ corresponds to the number of notes, while the columns include information on onset (in beats), duration (in beats), MIDI channel, MIDI pitch, velocity, onset (in seconds), and duration (in seconds). We will make
use of the given MIDI note numbers of the pitch as defined in the MIDI standard to identify the corresponding note letter name.  Further, we proceed by detecting simultaneously played notes, to do that we use the algorithm explained in Ref. \cite{mrad2024network}. The musical elements (notes, chords) are then put in order based on their existence in the composition. \\
The construction of the simplicial complexes involves three key contributions of musical elements for the purpose of putting together facets of both harmony and melody into a single cohesive representation. Consequently, we propose employing a representation that encodes both layered chords (notes played simultaneously) and transitions between notes over time. In other words, we are incorporating information of both vertical and horizontal  components of the music. The third aspect contributing to the construction is coming from the closure property of the simplicial complexes includes the proper subsets of the data (simplices).  To capture the vertical component, we deal with the harmonic structure as a simultaneity of notes and consider relationships between vertically stacked notes that are played at the same time.
 For the horizontal treatment, on the other hand, we look at the linear melodic aspect of the music, mainly the relationship between two consecutive notes or chords, which also preserves how the musical line and harmonic relationships evolve over time. In principle, we look at the sequence or progression of notes and chords and account for the correlation between them.\\
 The transitions between chords/notes are modeled based on the root notes of the chords. Specifically, we add an edge joining the root note of the chord with the root note of the neighboring chord in the case where we have a transition from a chord to another chord. In cases where single notes are involved, the edge joins the note itself with the following note or root of the following chord. We provide an illustrative example in Fig. \ref{musicSimplices}: A three-chord $G_4, B_4, D_5$ is nothing but a G major chord, the transition edge then joins $G$ to the next note in the sequence $F_4$. Similarly, the next transition connects $F_4$ to the root note of the chord formed by $C_4$ and $A_4$, which is an A minor chord, thus, it is an edge ($1$-simplex) given by $[F_4, A_4]$. \\
The components of the construction are as follows: notes and chords of different lengths, transitions, and their proper subsets. All of these span a simplicial complex of dimension $d$. For the case of violin music, the dimension of the simplicial complex is in general $d = 3$, which means that the simplex of highest order present in the simplicial complex is a $3-$simplex, or tetrahedron. In other words, it means we have, at most, chords of length four, and this constraint is coming from the nature of the instrument which only allows chords up to length four.

\subsection{Temporal evolution}
\begin{figure}[h]
    \centering
        \centering
        \includegraphics[width=.6\textwidth]{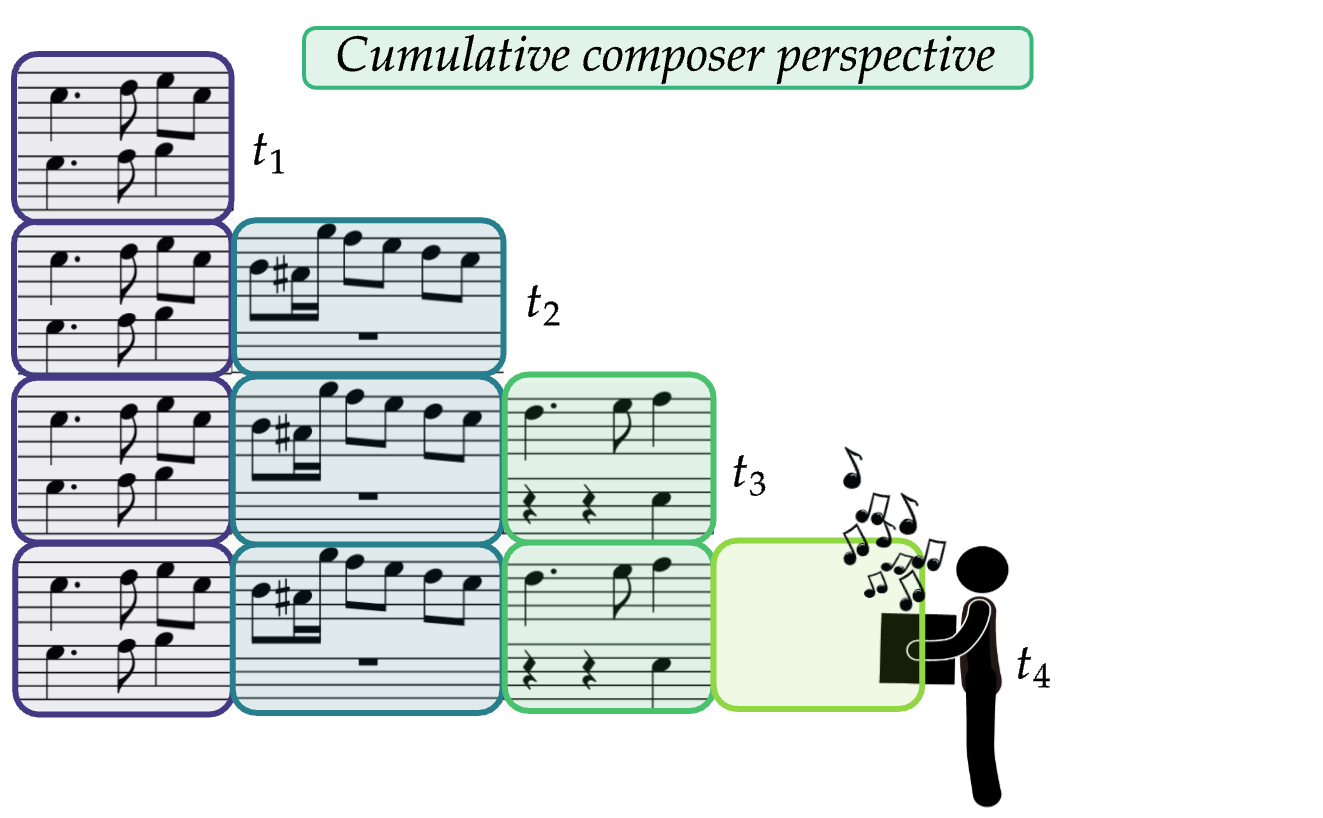}
        \caption{The scheme describing the cumulative approach}
        \label{cumulativeapproach}
\end{figure}

\begin{figure}[h]
    \centering
    \includegraphics[width=.8\linewidth]{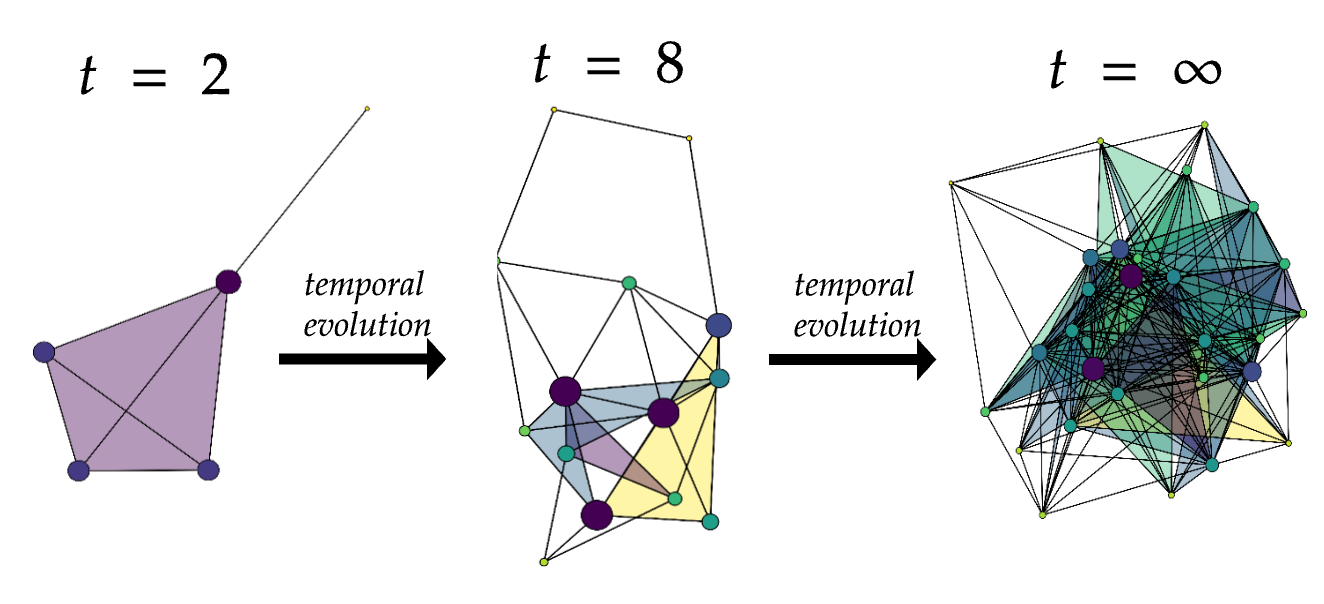}
    \caption{Temporal evolution of a simplicial complex associated with the Adagio from Bach's \textit{Sonata No. 1 in G Minor for Solo Violin}.}
    \label{simplexTempEv}
\end{figure}

In order to acquire a comprehensive understanding on how the musical piece evolves, we need to incorporate the temporal dimension into our analysis. Subsequently, we account for the  evolution of the music, by considering a cumulative approach that enables us to trace the sequence of musical elements as they develop all-throughout the composition. This involves constructing the musical piece starting with the first musical element and cumulatively adding others as time progresses until the full piece is formed. This permits the incremental tracking of the contribution of every new musical element to the overall structure of the music. We illustrate the process in Fig. \ref{cumulativeapproach} and an example of the cumulative evolution of a simplicial complex corresponding to a particular musical movement in Fig. \ref{simplexTempEv}.

\begin{figure}[h]
    \centering
        \centering
        \includegraphics[width=.6\textwidth]{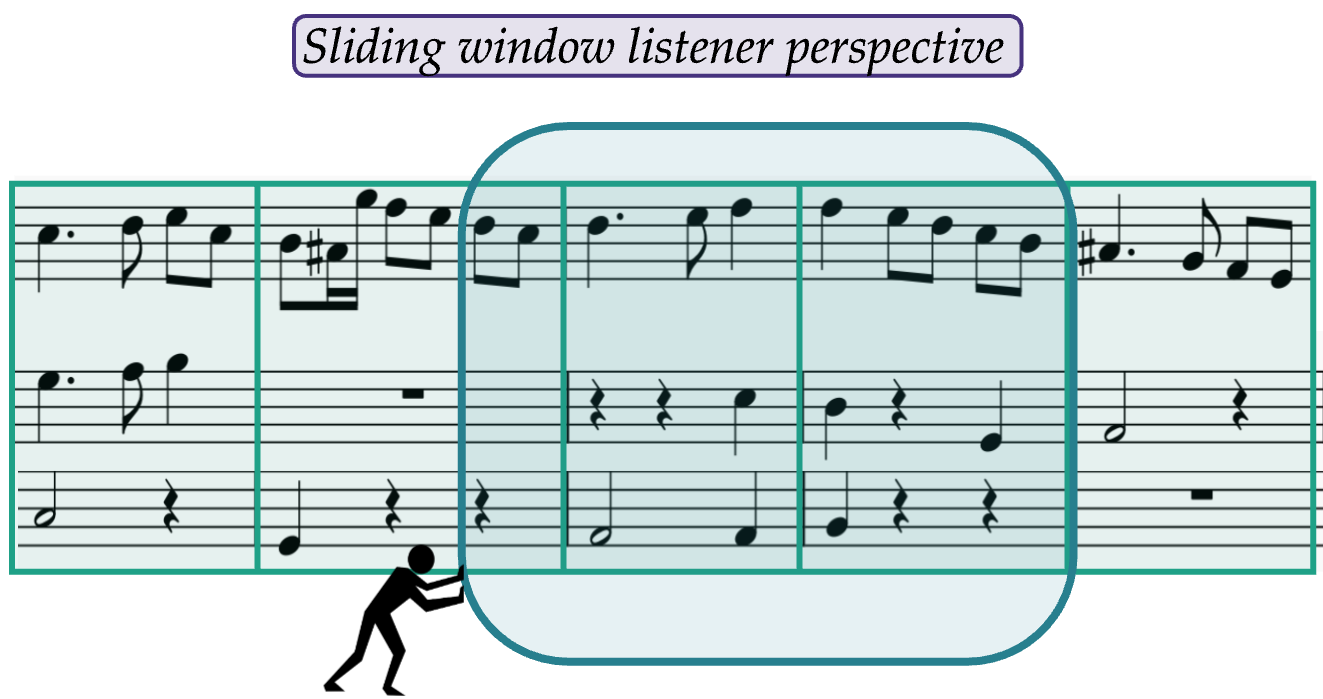}
        \caption{Figure illustrating the sliding window approach}
        \label{slidingwindow}
\end{figure}

The analysis, in some cases which will become clear later on, imposed a different kind of treatment: the sliding window technique. It consists of dividing the piece into segments that can be analyzed instantaneously as a listener would perceive it. We provide an illustration for the technique in Fig. \ref{slidingwindow}. We employ this approach because it provides information on how the musical elements behave structurally independent of the preceding or following components and gives expectations of the nuances a particular segment adds to the music. This approach, combined with the cumulative approach, were shown to be useful in the context of detecting musical form, and this will be thoroughly discussed later on as the results are presented. 
 
\begin{figure}[h]
     \centering
     \includegraphics[width=0.6\linewidth]{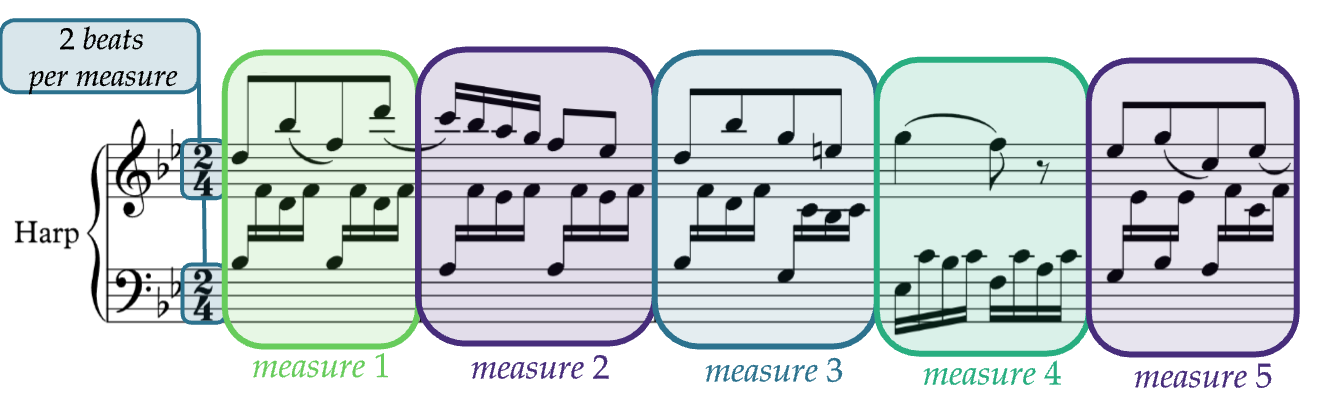}
     \caption{The scheme illustrating the process of segmentation of the musical piece}
     \label{segment}
 \end{figure}

Both of the temporal treatments presented above require practical segmentation schemes to segment the musical piece. The first step is selecting an appropriate segment size with a given number of musical elements per time step. It is musically intuitive to associate this with the musical measure as this reflects the underlying rhythmic structure of the piece. Since the MIDI files themselves do not inherently encode information about measure boundaries, it is then essential to begin by delimiting measures. We proceed by identifying the number of beats per measure determined from the meter, the process is illustrated in Fig. \ref{segment}

 \subsection{Analysis}
We use the processed musical data to construct the simplicial complexes according to the method outlined above and capture its temporal dynamics. Subsequently, we turn to the formalism described in the previous sections and begin by constructing the incidence matrices by operating the boundary maps on the simplices of all orders, like in Eq. \ref{EqIncidence}. For the $3-$dimensional simplicial complexes for the violin, we obtain three incidence matrices $\mathbf{B_1}$, $\mathbf{B_2}$, and $\mathbf{B_3}$. We then proceed by constructing the corresponding Hodge Laplacians as in Eq. \ref{eq:Laplacian} which serve as the fundamental operators for analyzing the topological structure of the system. We examine the spectral properties of these Hodge Laplacians, particularly, the dimension of their kernels and systematically track the evolution of topological invariants over time to understand how global connectivity properties evolve dynamically. Examining the temporal evolution of the Betti numbers would result in different Betti numbers of each order, all of which tend to increase due to the cumulative construction. It is more informative to instead track the evolution of the Euler characteristic, which combines all Betti numbers into a single topological invariant and is not expected to always increase, as expressed in Eq. \ref{eq:EulerChar}, owing to the alternating signs in its definition. \\

Following the topological analysis, we turn to the geometric characterization of the structures under study. In particular, we examine geometric properties of curvatures at vertices and edges by implementing the discrete Forman-Ricci curvature approach. In this context, the Gaussian curvature is defined over vertices, while the mean curvature is associated with edges. These curvature measures provide a local geometric description that complements the global topological invariants previously examined.\\
In the next stage of the analysis, we simultaneously examine the geometric and topological properties of the evolving structure by applying the discrete Gauss–Bonnet theorem. This theorem asserts a linear relationship between the sum of the local curvatures (computed via the Forman–Ricci curvature) and the global topological invariant, the Euler characteristic. We employ the theorem in our context to assess the extent to which the theorem manifests in our constructed structures across the different musical \textit{genres}: specifically, which movement adhere to the expected linear behavior, which movements deviate from it, and how the overall behavior evolves over time.

\section{Results}
\subsection{Topology}
\subsubsection{Slow movements}
\begin{figure}[h]
     \begin{minipage}{0.45\linewidth}
        \centering
        \includegraphics[width=\linewidth]{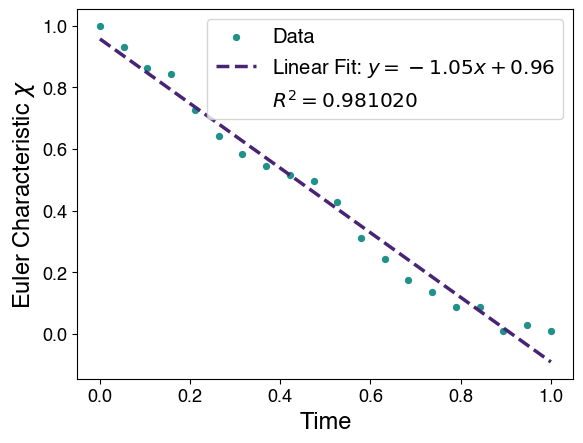}
        \caption{Linear fit of the normalized Euler characteristic of Adagio from sonata 1 with $R^2 = .98$}
        \label{fig:LinearFitEulerCharAdagio1}
    \end{minipage}
     \hfill
    \begin{minipage}{0.45\linewidth}
        \centering
        \includegraphics[width=\linewidth]{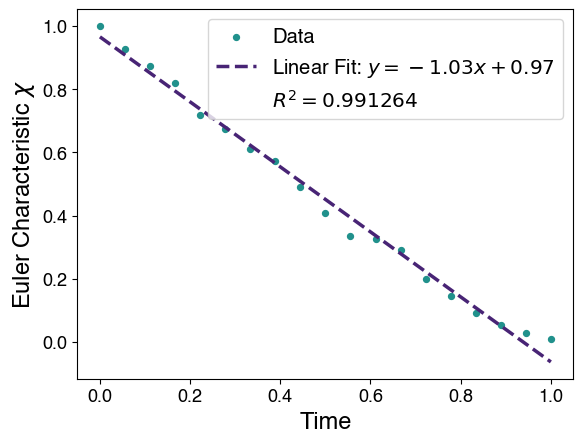}
        \caption{Linear fit of the normalized Euler characteristic of Grave from sonata 2 with $R^2 = .991$}
        \label{fig:FitEulerCharGrave2}
    \end{minipage}
\end{figure}

\begin{figure}[h]
   \begin{minipage}{0.45\linewidth}
        \centering
        \includegraphics[width=\linewidth]{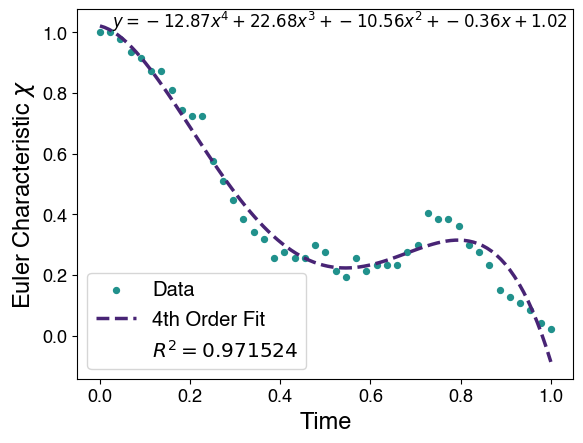}
        \caption{Polynomial of order $4$ fit given by $y = -12.87x^4 + 22.68x^3 + -10.56x^2 -0.36x + 1.02$ for the normalized Euler characteristic of Adagio from Sonata 3 with $R^2 =  .971$.}
        \label{fig:4thOrdFitEulerCharAdagio3}
    \end{minipage}
     \hfill
     \begin{minipage}{0.45\linewidth}
    \includegraphics[width=\linewidth]{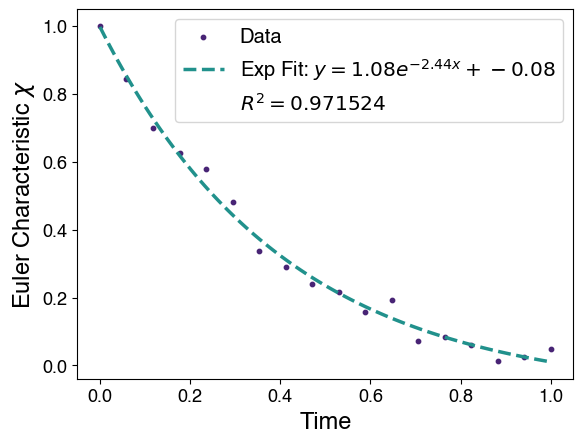}
        \caption{Exponential fit for the normalized Euler characteristic of Siciliano from Sonata 1 with exponent $\alpha = -2.44$ and $R^2 = .97$.}
        \label{fig:ExpFitEulerCharSiciliano1}
    \end{minipage}
\end{figure}

%For the movements we classified as \textit{slow} movements, 
We begin by analyzing the slow opening movements (also preceding the fugues) in Bach's solo violin sonatas, examining their Euler characteristic evolution. The Adagio from Sonata 1 and Grave (Sonata 2) exhibit nearly identical linear decays with slopes $-1.05$ and $-1.04$ respectively, sharing a common intercept at $y = 0.96$ (Figs.~\ref{fig:LinearFitEulerCharAdagio1}, \ref{fig:FitEulerCharGrave2}), suggesting similar structure in their gradual development. In contrast, Sonata 3's Adagio follows a fourth-order polynomial trajectory (Fig.~\ref{fig:4thOrdFitEulerCharAdagio3}), reflecting a more complex variation structure . The Siciliano following Sonata 1's fugue demonstrates exponential decay with exponent $-2.44$ (Fig.~\ref{fig:ExpFitEulerCharSiciliano1}). 

%%Sonata 1 Adagio: complete+ %%Sonata 2 grave: complete

 %% Sonata 3 adagio: complete + 
%%sonata 1 siciliano: complete

\subsubsection{Fugues}
\begin{figure}[!htp]
\begin{minipage}{0.48\linewidth}
        \centering
	\includegraphics[width=0.8\linewidth]{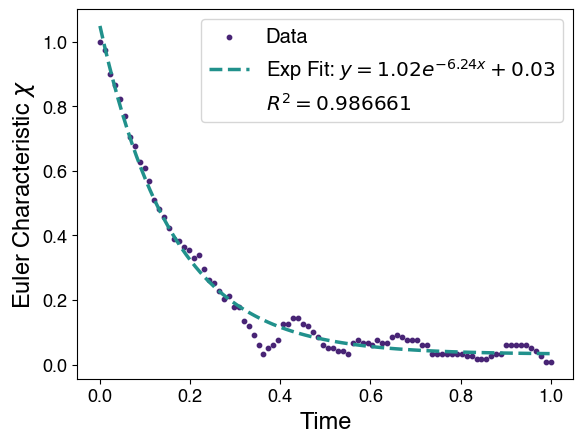}
        \caption{Exponential fit for the normalized Euler characteristic of Fugue from Sonata 1 with exponent $\alpha = -6.24$ and $R^2 = .986$.}
        \label{fig:FitEulerCharFugue1}
    \end{minipage} \hfill
\begin{minipage}{0.48\linewidth}
        \centering
        \includegraphics[width=0.8\linewidth]{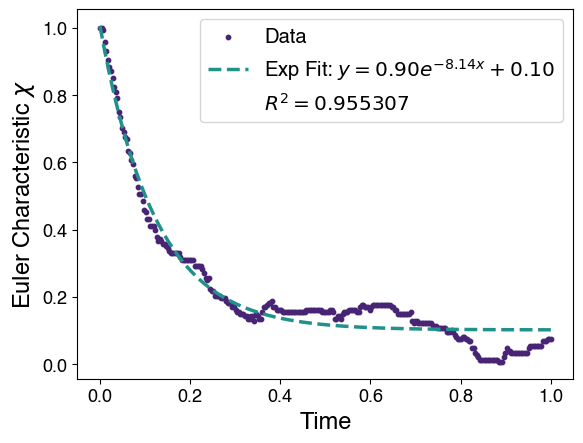}
        \caption{Exponential fit for the normalized normalized Euler characteristic of Fugue from Sonata 2 with exponent $\alpha = -8.14$ and $R^2$ = 0.955.}
        \label{fig:FitEulerCharFugue2}
        \end{minipage}
\end{figure}

The Fugues in our dataset, situated between two slow movements in their respective musical scores, generally exhibit an exponential trend in their Euler characteristic evolution over time. This is illustrated in Fig.~\ref{fig:FitEulerCharFugue1} (Fugue from Sonata~1, exponent: $-6.24$) and Fig.~\ref{fig:FitEulerCharFugue2} (Fugue from Sonata~2, exponent: $-8.14$).  
To further explore our results, we conduct a comparative analysis of solo violin (and solo string) Fugues by various composers, as well as Fugues for non-string instruments. Our case study examines fugues from diverse historical periods, including works by Max Reger, Bartolomeo Campagnoli, Vaclav Pichl, and Eugène Ysaÿe (see Appendix~\ref{appendix} for complete results). The analysis reveals that solo violin fugues consistently exhibit exponential decay in their Euler characteristic evolution across the repertoire. This exponential signature emerges as a stable topological feature that persists across different compositional styles (Baroque through more modern eras) and is hence robust to variations in technical complexity. The recurrence of this pattern suggests that the exponential decay in $\chi(t)$ represents a property of fugal writing for solo violin, transcending historical period and individual composers though the converse is not universally true. In contrast, Fugues for keyboard instruments, such as those from Bach's \textit{Well-Tempered Clavier} (Figs.~\ref{fig:BachWTC5Euler}--\ref{fig:BachWTC14Euler} in Appendix~\ref{appendix}), do not exhibit the same exponential behavior. This suggests a distinction in structural evolution between string Fugues and Fugues composed for different instruments (keyboard, for instance).

For Max Reger's solo violin Fugues, the Euler characteristic evolution in his Fugue in A minor (Fig. ~\ref{fig:MRegerAMinor}) and Fugue in D minor (Fig.~\ref{fig:MRegerDMinor} in Appendix \ref{appendix}) follows an exponential trend, with exponents of the form $-2.XX$. Similarly, Bartalomeo Campagnoli's Fugue in D minor (Fig. ~\ref{fig:BCampagnoliDMinor} in Appendix~\ref{appendix}) exhibits exponential decay with an exponent of $-2.72$. Eugène Ysaÿe's Fugue in G minor (Fig. ~\ref{fig:YsayeFugueGMinor} in Appendix~\ref{appendix}) also demonstrates an exponential decrease, with exponent $-2.21$. In the case of Vaclav Pichl, the exponents for his Fugues in D minor and D major (Figs. ~\ref{fig:VPichlDMinor} and ~\ref{fig:VPichlDMajor} in Appendix~\ref{appendix}) are $-1.84$ and $-2.18$, respectively.

\subsubsection{Dance movements}
\begin{figure}[!htp]
\centering
\begin{minipage}[b]{0.48\linewidth}
    \centering
    \includegraphics[width=0.8\linewidth]{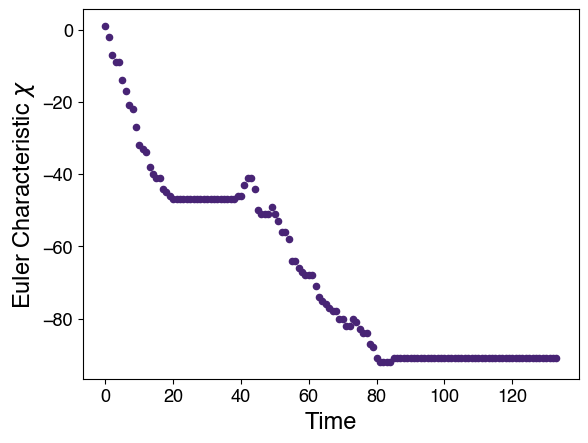}
    \caption{The evolution of the Euler characteristic of Bourr\'ee from partita 1}
    \label{fig:EulerCharBourree1}
\end{minipage}
\hfill
\begin{minipage}[b]{0.48\linewidth}
    \centering
    \includegraphics[width=0.8\linewidth]{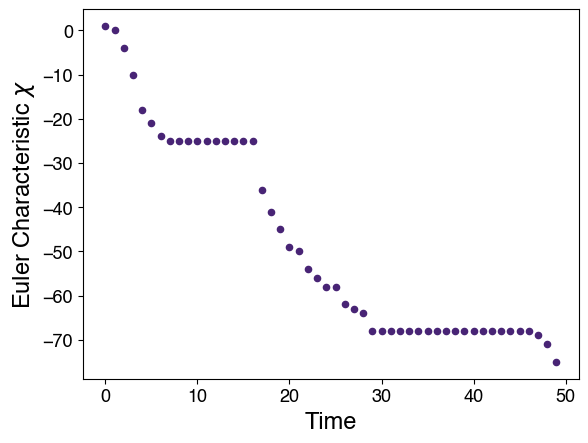}
    \caption{The evolution of the Euler characteristic of Sarabande from partita 2}
    \label{fig:EulerCharSarabande2}
\end{minipage}
\end{figure}

\begin{figure}[!htp]
        \centering
        \includegraphics[width=0.7\linewidth]{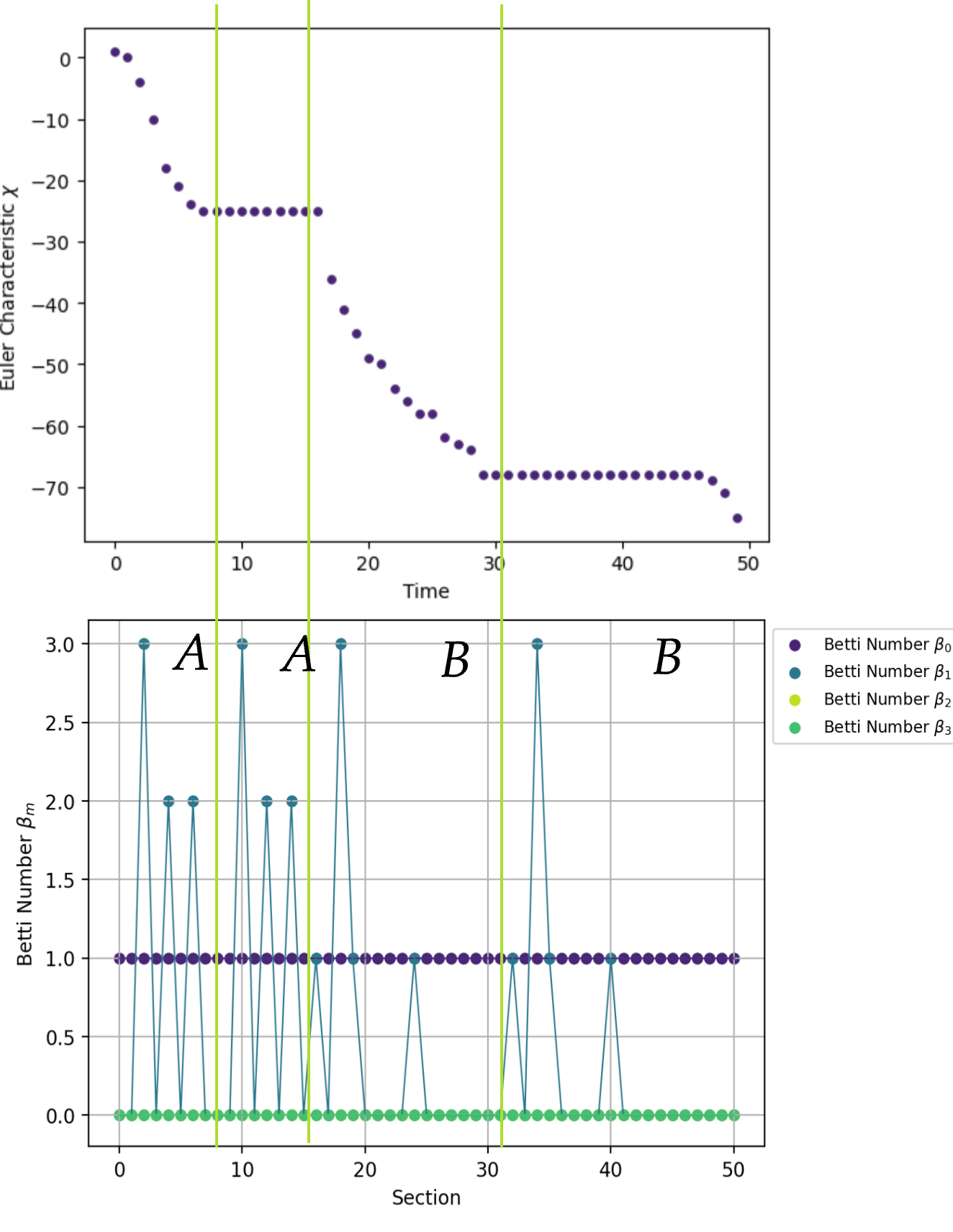}
        \caption{The sliding window technique applied to the Betti numbers of Sarabande from partita 2, in comparison with the cumulative approach result of the same movement.}
        \label{fig:FormSarabande2}
\end{figure}

All dance movements in our dataset are drawn from the partitas and exhibit binary form structure. In musical terms, binary form consists of two complementary sections (A and B), each typically repeated where A and B are two distinct themes. Section A establishes the tonic key and presents the main thematic material, and Section B provides contrasting new theme, often modulating to a different key before returning to the tonic. The Euler characteristic evolution in these movements reveals a distinctive pattern characterized by two prominent plateaus: one occurring approximately at the midpoint and another towards the end of the piece. This behavior is illustrated in Figs. ~\ref{fig:EulerCharBourree1}-\ref{fig:EulerCharSarabande2}.
To interpret these plateaus and identify the reason for their occurrence, we determine which structural changes in the simplicial complex correspond to the formation of plateaus in the Euler characteristic evolution by employing a sliding window analysis of the simplicial complex evolution to track incremental modifications in the simplicial complex structure at each temporal step and identify specific simplices that emerge or disappear during plateau regions. The results of the sliding window approach coupled with those of the cumulative approach for one of the movements (Sarabande from Partita 2) are presented in Fig. \ref{fig:FormSarabande2}. This reveals that plateaus in the evolution of the Euler characteristic correspond to repeated themes in the musical piece.

\clearpage
\subsection{Geometry}
\begin{figure}[!htp]
    \centering
    \begin{minipage}[b]{0.3\linewidth}
        \centering
        \includegraphics[width=\linewidth]{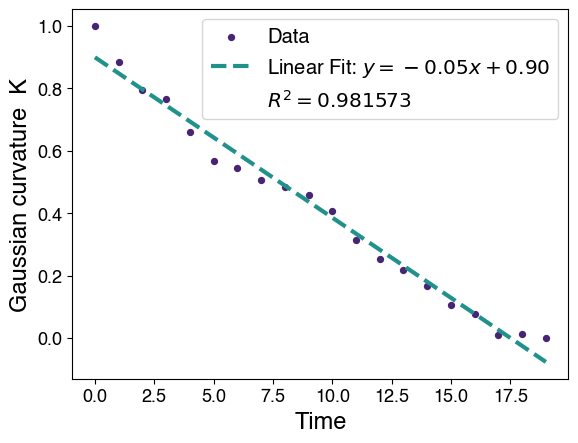}
        \caption{The evolution of the normalized Gaussian curvature as a function of time for Adagio from Sonata No.1.}
        \label{fig:Adagio1Order0}
    \end{minipage}
    \hfill
        \begin{minipage}[b]{0.3\linewidth}
        \centering
        \includegraphics[width=\linewidth]{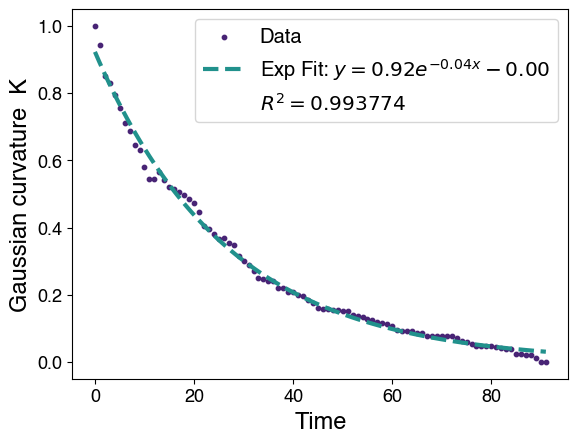}
        \caption{The evolution of the normalized Gaussian curvature as a function of time for Fugue from Sonata No.1.}
        \label{fig:Fugue1Order0}
    \end{minipage}
     \hfill
        \begin{minipage}[b]{0.3\linewidth}
        \centering
        \includegraphics[width=\linewidth]{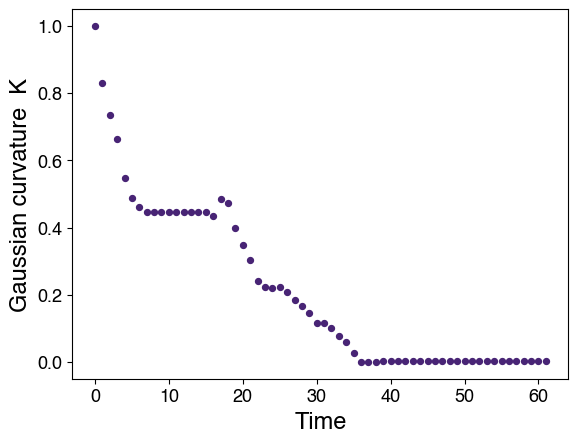}
        \caption{The evolution of the normalized Gaussian curvature as a function of time for Sarabande from Partita No.1.}
        \label{fig:Sarabande1Order0}
    \end{minipage}
\end{figure}

The second part of the analysis will deal with the geometrical elements associated with the simplicial complexes, in particular, measures of discrete curvature: the Forman-Ricci curvature. Each node, dyad, or hyperedge in the simplicial complex is assigned a curvature value. At each timestep \( t \), we evaluate the curvature of every simplex. We compute the Forman-Ricci curvature for each individual p-hyperedge, then determine the average p-hyperedge curvature by taking the mean curvature value across all p-hyperedges in the hypergraph.
Our primary focus, however, is not on static curvature distributions but on its dynamical evolution, as we proceeded in the previous section. We track the mean Forman-Ricci curvature over edges (analogous to Ricci tensor behavior in Riemannian geometry) and Gaussian curvature over vertices (analog of scalar curvature), observing how these quantities evolve over time. In general, the curvature trends mirror those observed for the Euler characteristic: linear patterns emerge for movements that exhibited linear Euler characteristics, exponential decays match those with exponential Euler characteristic behavior, and plateaus appear corresponding to repetitive sections. To demonstrate the correspondence, we provide three representative examples in Figs.~\ref{fig:Adagio1Order0}, ~\ref{fig:Fugue1Order0} and  ~\ref{fig:Sarabande1Order0} one for each case (linear, exponential, and plateau regimes).

%%Adagio1
% \clearpage
\subsection{Gauss-Bonnet theorem}
\begin{figure}[!htp]
    \centering
    \includegraphics[width=0.95\linewidth]{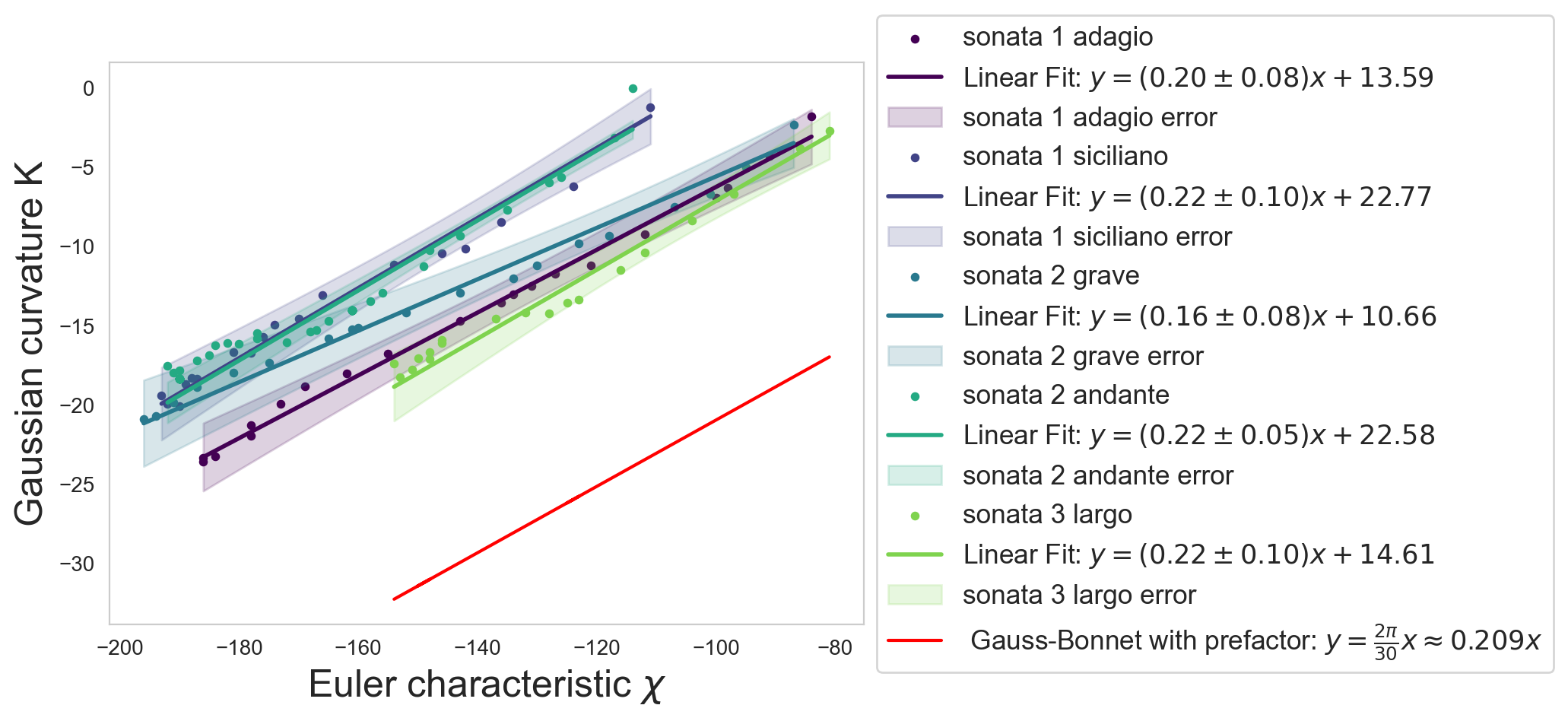}
    \caption{Verification of the Gauss-Bonnet theorem for simplicial complexes of movements}
    \label{verifyGauss}
\end{figure}

Having evaluated both topological and geometric measures, we now validate the Gauss-Bonnet theorem in our musical context through the relationship between Euler characteristic $\chi$ and Gaussian curvature $K$. Our results reveal distinct patterns between movement types. Slow movements maintain the expected linear relationship throughout and are closest to the theoretical prediction. This demonstrates the robust validity of the Gauss-Bonnet theorem for slow movements, regardless of their Euler characteristic ($\chi$) evolution patterns. We compare the theoretical prediction of the Gauss-Bonnet theorem with slope $2\pi$ to the empirical slopes of the linear fits recovered from the musical data. The calculated slopes from the musical movements fall within a narrow window in the vicinity of $0.2$ as displayed in Fig. \ref{verifyGauss}. 
We find that by introducing a normalization prefactor of $\alpha \approx \frac{1}{30}$, we recover the theoretical expectation. The scaling factor arises naturally in the context of the discrete nature of the simplicial complex representation: the complex size, given by the number of nodes (ranging between 26 and 32 for our chosen musical movements) serves as the appropriate normalization constant. The rescaling yields the relationship $\alpha \sum_v K_v \approx 2\pi\chi$ where $K_v$ denotes the curvature at vertex $v$. This demonstrates that when properly normalized by system size, the discrete simplicial complexes representing the musical data of slow movements obey the Gauss-Bonnet theorem, with the total curvature summing to $2\pi\chi$ as expected in the continuous limit.
Fugues, on the other hand, exhibit a different behavior, with initial instability manifesting as pronounced zigzag patterns before stabilizing into linearity, results are moved to Appendix~\ref{appendix} in Fig. \ref{fig:Fugue1PlotEulerAverageRicciCurvature}. Finally, Dance movements follow the linear trend initially but deviate at a later stage, the results are also moved to Appendix~\ref{appendix}in Fig. \ref{fig:Bourree1PlotEulerAverageRicciCurvature}. A single sample of the slow movements results is also in the Appendix~\ref{appendix} in Fig. \ref{fig:Adagio1PlotEulerAverageRicciCurvature}.

\newpage
\section{Discussion}
In this paper, we borrowed notions from topology and discrete geometry and employed them in the context higher-order networks, particularly, simplicial complexes. 
We examined topological invariants evolving over time for the different musical movements which we classified into three categories: slow movements, fugues, and dance movements. The results demonstrated that slow movements do not conform to a characteristic pattern and exhibit heterogeneous evolution patterns. This spectrum of behaviors: linear, polynomial, exponential, etc... reveals that there is no topological signature that characterizes the slow movements in this repertoire. Rather, their Euler characteristic evolution appears to be determined by movement-specific compositional strategies. For instance, for some movements (Adagio from Sonata No.1 and Grave from Sonata No.2), the Euler characteristic decays linearly. The linear behavior could be attributed to the long sequences of eighth notes, which are interrupted by a change in chords at regular intervals; the latter creating predictable changes in the shape of the complex. Moreover, these movements include re-statements of the original theme in a modulated key, which means that similar simplices that do not overlap with previously created structures are generated in the simplicial complexes as time evolves, possibly yielding the linear trend in how the topology evolves. Other slow movements have different structural organizations and compositional strategies, which produce distinct trends, such as the Adagio from Sonata No.3, which takes the form of a fourth-order polynomial, and the Siciliano, whose Euler characteristic evolves as an exponential decay. Interestingly, Fugues consistently showed exponentially decaying behavior in their Euler characteristic, and this feature was found across composers and eras. What most characterizes fugues is their use of counterpoint, a feature they share with the Siciliano, which features a distinctive three-part contrapuntal writing consisting of one bass and two soprano voices. The fact that exponential behavior appears only in solo violin fugues suggests that the nature of the string instrument itself might constrain how the topology evolves. The plateaus characterizing the evolution of the Euler characteristic of the Dance movements arise from the repetitive formal structure of the piece. During the repetitions of the musical themes, the simplices generated are identical copies of previously constructed structures in the complex. Mathematically, this means that the change in the Euler characteristic $\chi$ is zero because the newly added simplices, already taking part in the simplicial complex are do not change the topology, until new musical structures disrupt the repetition.
As for the geometry analysis, we coupled it with the topological feature to explore the adherence of the musical data to the Gauss-Bonnet theorem. Our findings suggest that slow movements conform perfectly to the predicted theoretical expectation of the Gauss-Bonnet theorem, with a normalization prefactor proportional to the size of the simplicial complex, while other compositional genres (Fugues, Dance movements...) do not. For Fugues, we observe jagged lines for the first few measures, which likely correspond to the initial phase of the Fugue: the exposition, where each voice presents the melodic subject in turn. The subsequent \textit{stabilization} into linearity is likely attributed to the instant where all voices have entered, and the music reaches a stable level of complexity in the development phase. For dance movements, the deviation from linearity happens towards the end, around the restatement of the second theme (second B in ABAB form). The precise reason after the perfect adherence to the Gauss-Bonnet theorem in slow movements remains to be fully characterized and compared with the movements which do not, particularly the exposition section of Fugues and repetition of second theme in Dance movements.

\section{Conclusion}
This work presents a topological framework for the purpose of analyzing J. S. Bach's Solo Violin Sonatas and Partitas through simplicial complexes employed as higher-order networks. This representation encodes musical transitions and elements more faithfully than traditional networks, capturing higher-order structures and interactions such as chords and simultaneously played notes. Although this approach introduces connections that are not explicitly present in the original dataset (the musical score), but emerge from the closure property of the simplicial complexes, it captures genre-specific features through topological metrics. The complex polyphonic structures in this analysis are distilled into low-dimensional projections, sacrificing musical details while preserving distinctive traits of certain compositional genres in the evolution of their Euler characteristic and discrete geometric curvature analysis. The geometric connection through curvature, though exploratory, reveals interesting correlations between compositional strategies and spatial properties of Bach's works. This perspective opens up new directions for quantitatively studying musical structures through algebraic topology and geometry, despite the inherent limitations stemming from the dimensionality reduction of the simplicial complex representation to one-dimensional scalar metrics (Euler characteristic and curvature). Future work may refine our methods by incorporating the directed element in the representation and further investigating the connection to geometry.

\section*{Acknowledgments} D.M received Research Assistantship (RA) support for this work from the Center for Advanced Mathematical Sciences (CAMS) at the American University of Beirut. S.N and D.M are grateful for the discussions with Prof. Jihad Touma, CAMS's Director, Khaled Yassine, CAMS's Musician in Residence, and Prof. Pablo Padilla. 

\clearpage
 \bibliographystyle{elsarticle-num} 
 \bibliography{cas-refs}

 \newpage
\appendix
\renewcommand{\thesection}{\Alph{section}}
\renewcommand{\thetable}{\thesection\arabic{table}}
\renewcommand{\thefigure}{\thesection\arabic{figure}} 

\setcounter{section}{0} 
\setcounter{table}{0}
\setcounter{figure}{0}

\section{\large Appendix} \label{appendix} 
\addcontentsline{toc}{section}{Appendix} Below, we present the supplementary results that support the findings discussed above.

\begin{figure}[h!]
    \centering
    \begin{minipage}[b]{0.48\linewidth}
        \centering
        \includegraphics[width=\linewidth]{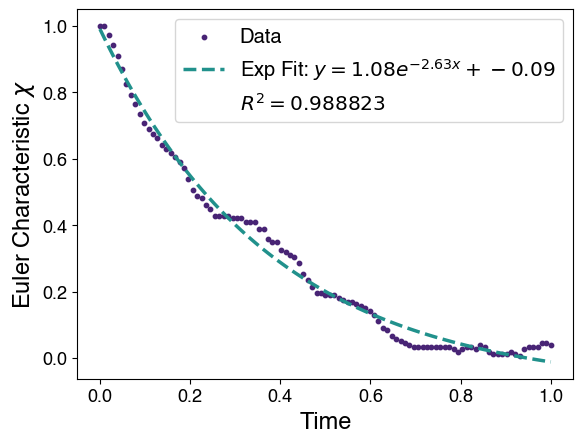}
        \caption{Evolution and exponential fit of the normalized Euler characteristic for Max Reger, Fugue, A Minor}
        \label{fig:MRegerAMinor}
    \end{minipage}
    \hfill
    \begin{minipage}[b]{0.48\linewidth}
        \centering
        \includegraphics[width=\linewidth]{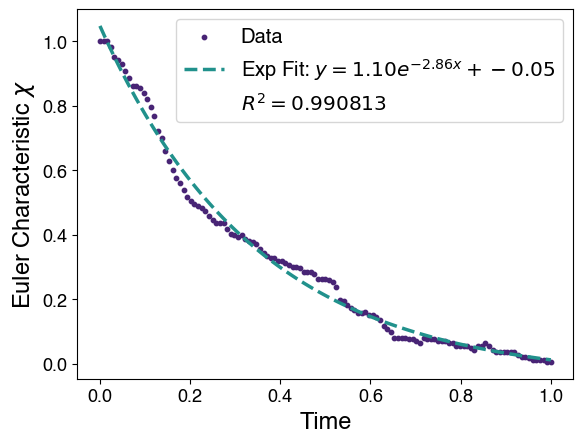}
        \caption{Evolution and exponential fit of the normalized Euler characteristic for Max Reger, Fugue, D Minor}
        \label{fig:MRegerDMinor}
    \end{minipage}
\end{figure}

\begin{figure}[h!]
    \centering
   \begin{minipage}[b]{0.48\linewidth}
        \centering
        \includegraphics[width=\linewidth]{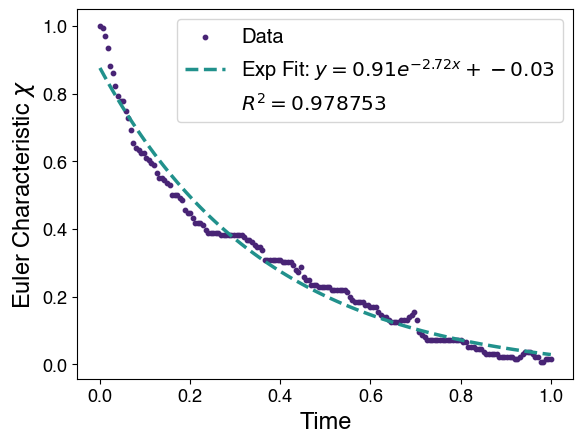}
        \caption{Evolution and exponential fit of the normalized Euler characteristic for Bartolomeo Campagnoli, Fugue, D Minor}
        \label{fig:BCampagnoliDMinor}
    \end{minipage}
    \hfill
    \begin{minipage}[b]{0.48\linewidth}
        \centering
        \includegraphics[width=\linewidth]{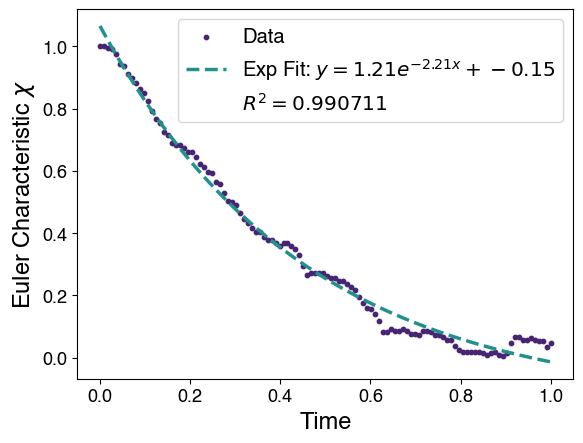}
        \caption{Evolution and exponential fit of the normalized Euler characteristic for Ysaÿe, Fugue, G Minor}
        \label{fig:YsayeFugueGMinor}
    \end{minipage}
\end{figure}

\begin{figure}[h!]
    \centering
    \begin{minipage}[b]{0.48\linewidth}
        \centering
        \includegraphics[width=\linewidth]{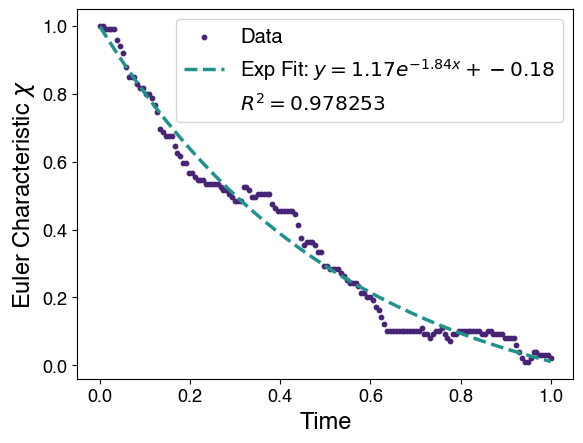}
        \caption{Evolution and exponential fit of the normalized Euler characteristic for Vaclav Pichl, Fugue, D Minor}
        \label{fig:VPichlDMinor}
    \end{minipage}
    \hfill
    \begin{minipage}[b]{0.48\linewidth}
        \centering
        \includegraphics[width=\linewidth]{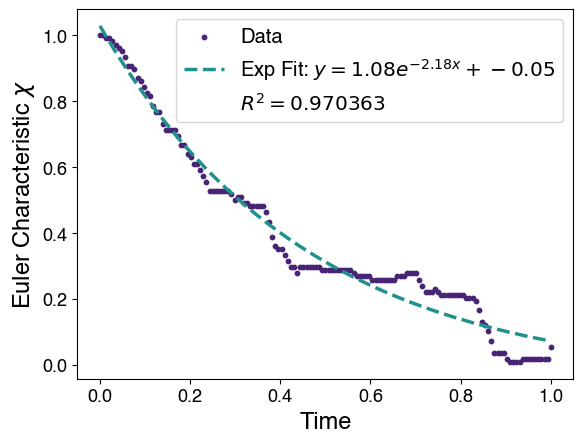}
        \caption{Evolution and exponential fit of the normalized Euler characteristic for Vaclav Pichl, Fugue, D Major}
        \label{fig:VPichlDMajor}
    \end{minipage}
\end{figure}

\begin{figure}[h!]
    \centering
     \begin{minipage}[b]{0.315\linewidth}
        \centering
        \includegraphics[width=\linewidth]{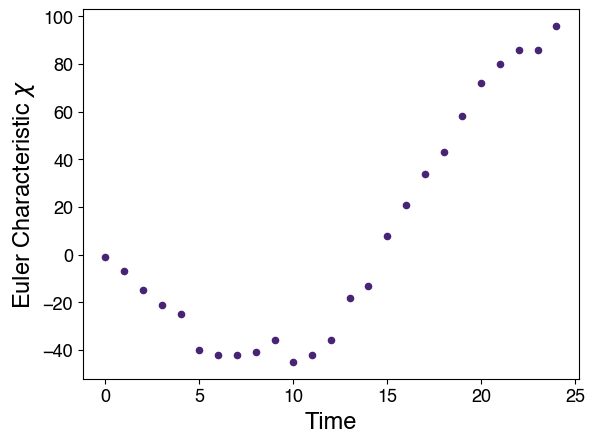}
        \caption{Evolution of the normalized Euler characteristic for Bach, WTC Fugue 5}
        \label{fig:BachWTC5Euler}
    \end{minipage}
    \hfill
    \begin{minipage}[b]{0.315\linewidth}
        \centering
        \includegraphics[width=\linewidth]{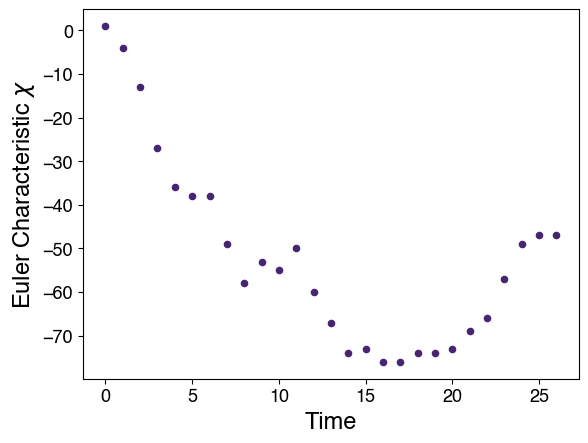}
        \caption{Evolution of the normalized Euler characteristic for Bach, WTC Fugue 9}
        \label{fig:BachWTC9Euler}
    \end{minipage}
    \hfill
    \begin{minipage}[b]{0.315\linewidth}
        \centering
        \includegraphics[width=\linewidth]{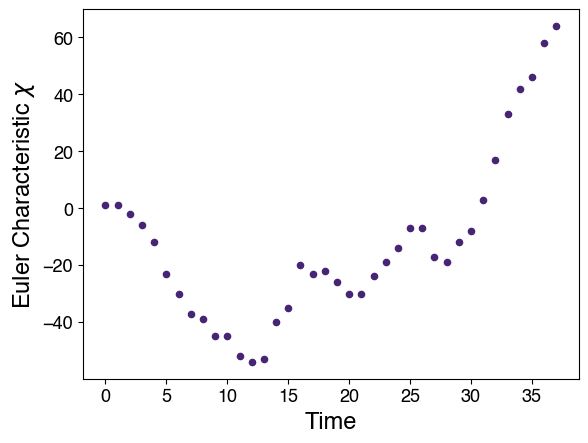}
        \caption{Evolution of the normalized Euler characteristic for Bach, WTC Fugue 14}
        \label{fig:BachWTC14Euler}
    \end{minipage}
\end{figure}

\begin{figure}[h]
    \centering
    \begin{minipage}[b]{0.32\linewidth}
        \centering
        \includegraphics[width=\linewidth]{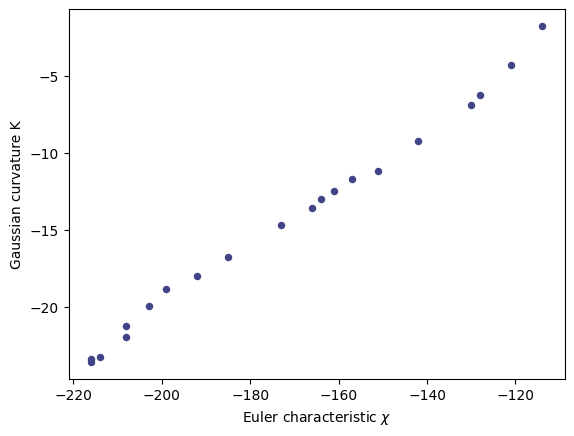}
        \caption{Evolution of Gaussian curvature as a function of Euler characteristic for Adagio from Sonata No. 1.}
        \label{fig:Adagio1PlotEulerAverageRicciCurvature}
    \end{minipage}
    \hfill
    \begin{minipage}[b]{0.32\linewidth}
        \centering
        \includegraphics[width=\linewidth]{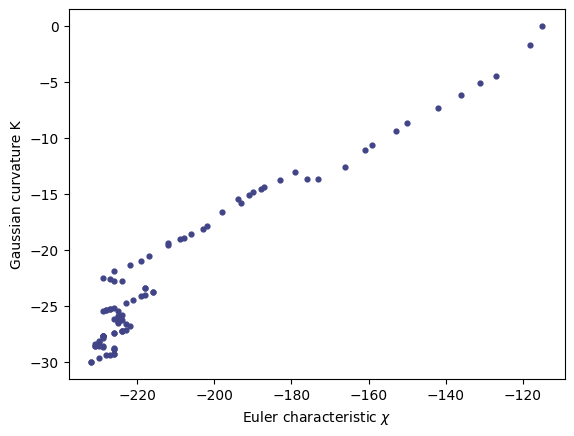}
        \caption{Evolution of Gaussian curvature as a function of Euler characteristic for Fugue from Sonata No. 1 .}
        \label{fig:Fugue1PlotEulerAverageRicciCurvature}
    \end{minipage}
	\hfill
    \begin{minipage}[b]{0.32\linewidth}
        \centering
        \includegraphics[width=\linewidth]{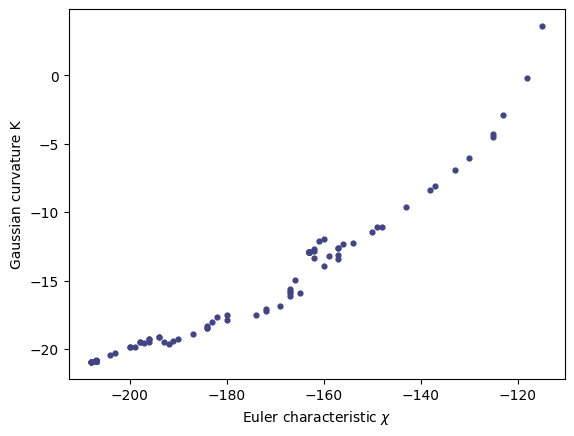}
        \caption{Evolution of Gaussian curvature as a function of Euler characteristic for Bourr\'ee from Partita No. 1.}
        \label{fig:Bourree1PlotEulerAverageRicciCurvature}
    \end{minipage}
\end{figure}

\end{document}